\documentclass[aps,prx,reprint,superscriptaddress, longbibliography]{revtex4-1}
\usepackage{H1}
\usepackage{amsmath}
\usepackage[pdftex,colorlinks=true]{hyperref}
\hypersetup{
citecolor = blue
}
\usepackage[normalem]{ulem}
\usepackage{amssymb}
\usepackage{amsthm}
\usepackage{amsfonts}
\usepackage{bbm}
\usepackage{array}

\begin{document}

\title{Localization from Hilbert space shattering: from theory to physical realizations}
\author{Vedika Khemani}
\affiliation{Department of Physics, Stanford University, Stanford, CA 94305, USA}
\author{Michael Hermele}
\affiliation{Department of Physics and Center for Theory of Quantum Matter, University of Colorado, Boulder, CO 80309}
\author{Rahul Nandkishore}
\affiliation{Department of Physics and Center for Theory of Quantum Matter, University of Colorado, Boulder, CO 80309}

\begin{abstract} 
We show how a finite number of conservation laws can globally `shatter' Hilbert space into exponentially many dynamically disconnected subsectors, leading to an unexpected dynamics with features reminiscent of both many body localization and quantum scars. A crisp example of this phenomenon is provided by a `fractonic' model of quantum dynamics constrained to conserve both charge and dipole moment. We show how the Hilbert space of the fractonic model dynamically fractures into disconnected \emph{emergent} subsectors \emph{within} a particular charge and dipole symmetry sector.  This shattering can occur in arbitrary spatial dimensions. A large number of the emergent subsectors, exponentially many in system volume, have dimension one and exhibit strictly localized quantum dynamics---even in the absence of spatial disorder and in the presence of temporal noise.  Other emergent subsectors display non-trivial dynamics and may be constructed by embedding finite sized non-trivial blocks into the localized subspace. While `fractonic' models provide a particularly clean realization, the shattering phenomenon  is more general, as we discuss.  We also discuss how the key phenomena may be readily observed in near term ultracold atom experiments. In experimental realizations, the conservation laws are approximate rather than exact, so the localization only survives up to a prethermal timescale that we estimate. We comment on the implications of these results for recent predictions of Bloch/Stark many-body localization. 
\end{abstract}
\maketitle

\normalsize
\section{Introduction}
 A particularly interesting question in many body quantum dynamics is whether a system can robustly fail to come to equilibrium under its own dynamics. One well known class of problems where such robust ergodicity breaking does arise involves the phenomenon of many body localization (MBL)~\cite{Anderson58, BAA, GMP, PalHuse, OganesyanHuse, Prelovsek, Imbrie2016, mblarcmp, mblrmp}. The lack of ergodicity in MBL systems follows from the presence of extensively many \emph{emergent} local integrals of motion~\cite{Huse14, Serbyn13cons, Imbrie2016}. 
 Other phenomena involving ergodicity breaking include integrable systems which also possess an extensive number of conserved quantities and, more recently, systems exhibiting so called `quantum scars' in which a vanishing fraction of eigenstates are non-thermal and coexist with thermal eigenstates~\cite{ShiraishiMori, AKLT1,  turner2018weak}. These scarred systems violate a `strong' form of the eigenstate thermalization hypothesis (ETH)~\cite{Deutsch,Srednicki, Rigol} which requires \emph{all} many-body eigenstates to individually look thermal; the weak form of the ETH, in which only almost all eigenstates are thermal, is known to not be sufficient to guarantee thermalization \cite{Kollath}. In these latter contexts of integrable and scarred models, however, it is not known to what extent the ergodicity breaking is robust to generic perturbations of the Hamiltonian~\cite{ck}, and explanations for the phenomenology of scarring are still being widely debated~\cite{ShiraishiMori, ck, ChoiAbanin, TDVPScars, lerose, ichinose, konik, Iadecola, IadecolaFH, paifractons_confinement}.  The search for alternative mechanisms for {\it robustly} and {\it provably }breaking ergodicity therefore continues apace. 

In this work, we introduce a novel mechanism for ergodocity breaking by which a \emph{finite} $O(1)$ number of conservation laws can \emph{provably} give rise to a dramatic fracturing of Hilbert space into {\it exponentially} many dynamical subsectors --- whence the word ``shatter" --- so that states even with the same \emph{global} quantum numbers for the conservation laws cannot mix under \emph{local} dynamics. This mechanism for localization is robust, even to temporal noise (unlike MBL), works in arbitrary spatial dimensions, and lies outside the framework of locator expansions. 

Most of this work will focus on models inspired by fracton systems~\cite{chamon, haah, fracton1, fracton2, sub, genem, fractonarcmp} in which excitations are known to exhibit restricted mobility. While much work on fractons has focused on exactly solvable spin models in 3D (which realize gapped fractonic phases), a useful complementary perspective on gapless fracton phases is provided by `higher-rank' gauge theories that conserve not only a $U(1)$ charge, but also higher multipoles of charge~\cite{sub}.  Motivated by this, Ref.~\cite{pai2018localization} considered a model of local random unitary circuit dynamics \cite{Nahum1, Nahum2, Nahum3,  KhemaniVishHuse,  Keyserlingk1, Keyserlingk2, Nahum4} constrained to conserve both a $U(1)$ charge and its dipole moment in one dimension, but with no other constraints. 
The mixed state dynamics of operators in this circuit showed signatures of localization for unitary circuits with range three interactions (but not longer ranged interactions), but the result was left largely unexplained~\cite{erratum}. 

The mechanism for localization via shattering that we introduce herein rigorously explains the results of \cite{pai2018localization} as a special case, but has far broader applicability. Indeed, one of our main results is an analytic \emph{proof} that the conservation of charge and dipole moment, along with spatial locality, is sufficient to produce exponentially many strictly `inert' states which live in dynamical subspaces of dimension exactly exactly equal to one and are left invariant by the dynamics. These look like simple product states in the computational basis, and have zero entanglement. 
This behavior is particularly striking since conventional wisdom holds that the presence of a finite ($O(1)$) number of commuting conservation laws should \emph{not} generically impede thermalization -- which requires instead the presence of extensively many explicit or emergent integrals of motion. Much as the case of MBL, the existence of these localized states could have interesting applications for protecting quantum information, for example in building quantum memories which remember their initial conditions and do not decohere. 
More generally, the shattering leads to a wide distribution of dimensions for the emergent subsectors, leading to a strong initial state dependence in the dynamics. We also explain how the dynamics of pure states can look localized for any finite ranged fractonic model (unlike mixed state dynamics which were considered in Ref.~\cite{pai2018localization} and are only localized for models with spatial range less than three).

The shattering of Hilbert space is robust in that it only relies on spatial locality and two local commuting constraints, and does not depend on details of the Hamiltonian, nor on the presence or absence of spatial or temporal translation symmetry. This is, again, quite striking since MBL systems are \emph{not} robust to temporal noise, and require spatial non-uniformity either in the form of random or quasiperiodic couplings. 
While the fractonic circuit provides an especially clean realization of localization from shattering, the phenomenon is more general and we also discuss alternate settings in which such dynamics may arise. In particular, we explain how the phenomenon may be rigorously realized in {\it arbitrary} spatial dimensions on hypercubic lattices, and also how the resulting phenomenology may be accessed in near term ultracold atom experiments. We apply this understanding to the special case of ultra cold atoms in a tilted potential, a problem with a long history \cite{Wannier, Greiner, SubirGirvin, FlachStark, RoschTilted, WernerDampedBloch}, which has recently been revisited~\cite{refaelbloch, pollmannstark, BakrTilted}, especially from the point of view of `Bloch' \cite{refaelbloch} or `Stark' \cite{pollmannstark} many body localization.

We note that physics analogous to Hilbert space fracture has also been observed in other models including, e.g., the Fermi-Hubbard model and its cousins~\cite{BernevigFH, IadecolaFH},  models with kinetic constraints (including in classical settings)~ \cite{LanGarrahan, OlmosLesanovsky, Gopalakrishnan_automata},  and dimer models~\cite{dimerfracture}. However, although constraints can lead to disconnected subsectors of Hilbert space in these cases, there is no understanding of general conditions that lead to Hilbert space fracture in the absence of fine tuning, and there is moreover no principled way to examine the stability of fracturing in these models to the addition of perturbations or noise. For example, the simplest kinetically constrained models comprise spin 1/2 systems in which the spin on a site can flip if certain conditions are obeyed by its neighbors, for instance if both neighbors are down. However, there is no unique or natural way to ``extend" such models, for example, to include the effect of further neighbor spins. Likewise, the dynamics in quantum dimer models come from certain ``flippable" plaquettes which are lattice dependent~\cite{QDM_review}. While allowing for longer flippable loops decreases fracture~\cite{dimerfracture}, there are no general results on how the number of disconnected sectors scales with such perturbations. In contrast, our work furnishes a robust class of constrained models where such Hilbert space fracture can be proven to exist on very general grounds. Moreover, fracture in our models comes from a clear physical origin --- the conservation of charge and dipole moment --- which furnishes a natural class of symmetry respecting perturbations, and also allows a natural generalization of the results to systems with longer range terms in the Hamiltonian, or to systems in higher dimensions. 

This work is organized as follows: we begin in Section \ref{sec: model} by introducing a simple (but not fine tuned) model in one dimension which realizes a shattered Hilbert space. We analyze this model in Section \ref{sec: analyze} and rigorously prove shattering. In Section \ref{sec: HigherD} we demonstrate how the phenomenology may be extended to systems in arbitrary space dimensions. In Section \ref{sec: nonfractonic} we discuss more general classes of quantum dynamics that exhibit a shattered Hilbert space. Near term physical realizations are discussed in Section \ref{sec: physical}, following which we conclude in Section \ref{sec: discussion} with a discussion of the implications of our results and some open directions. The appendices contain details of parenthetical importance to the main narrative. 

\section{The one dimensional model}
\label{sec: model}

Throughout this paper we will restrict to systems on hypercubic lattices with linear extent $L$ in each direction \emph{i.e.} one dimensional systems, square lattices in two dimensions, or simple cubic lattices in three dimensions. On each site $\mathbf{r}$, there exists an effective local $U(1)$ `charge'. This could be particle number for a bosonic or fermionic model, or $S^z$ (the $z$ component of spin) in a model of qudits with spin $S$. We will work with spin variables in most of what follows, but our statements are readily translated to the bosonic and fermionic cases. The dynamics will be required to conserve the total $U(1)$ charge ($Q= \sum_{\mathbf{r} }S_\mathbf{r}^z$), and certain multipole moments of charge (defined below in the obvious manner). 

We further assume that the dynamics are generated by strictly local models, such as static Hamiltonians with interactions of maximum spatial range $\ell$ or, more generally, models of unitary circuits with local gates that may be chosen randomly in space and time (and no gate acts on two sites separated by more than $\ell$ along any lattice axis). Our arguments are cleanest for a finite $O(1)$ interaction range $\ell$, but we also discuss exponentially local (rather than strictly local) models in the section on physical realizations. It will not matter whether our models are translationally invariant in space or time. 

We begin by analyzing the one dimensional model of quantum circuit dynamics introduced in \cite{pai2018localization}. The Hilbert space consists of a chain of $S=1$ quantum spins of length $L$, acted upon by local unitary gates which locally conserve both charge ($Q = \sum_j S_j^z$) and dipole moment ($P = \sum_{ j} j S^z_j,$), where $j$ is a site label. We can work with basis states in the $S^z$ basis, as these are eigenstates of both $P$ and $Q$.  On each site, the allowed values of $S^z$ are $|+\rangle, |-\rangle, |0\rangle$. The twin conservation laws greatly restrict the allowed movement of charges (fractons), as is characteristic of fracton phases \cite{sub, fractonarcmp}. For example, a single $+$ or $-$ charge on site $r$ has dipole moment $P= \pm r$. Such a charge cannot simply ``hop" to the left or right, because such a movement changes the net dipole moment by one unit. On the other hand, bound states of charges  or ``dipoles" of the form $(-+)$ have net charge zero and net dipole moment $P=\pm1$ independent of position, and these can move freely through the chain. Additionally, dipoles can enable the movement of charges, because a charge can move if it simultaneously emits a dipole to keep $P$ unchanged: $|0+0\rangle \rightarrow |+-+\rangle$.

\begin{figure}[t]
\centering
 \includegraphics[width=\columnwidth]{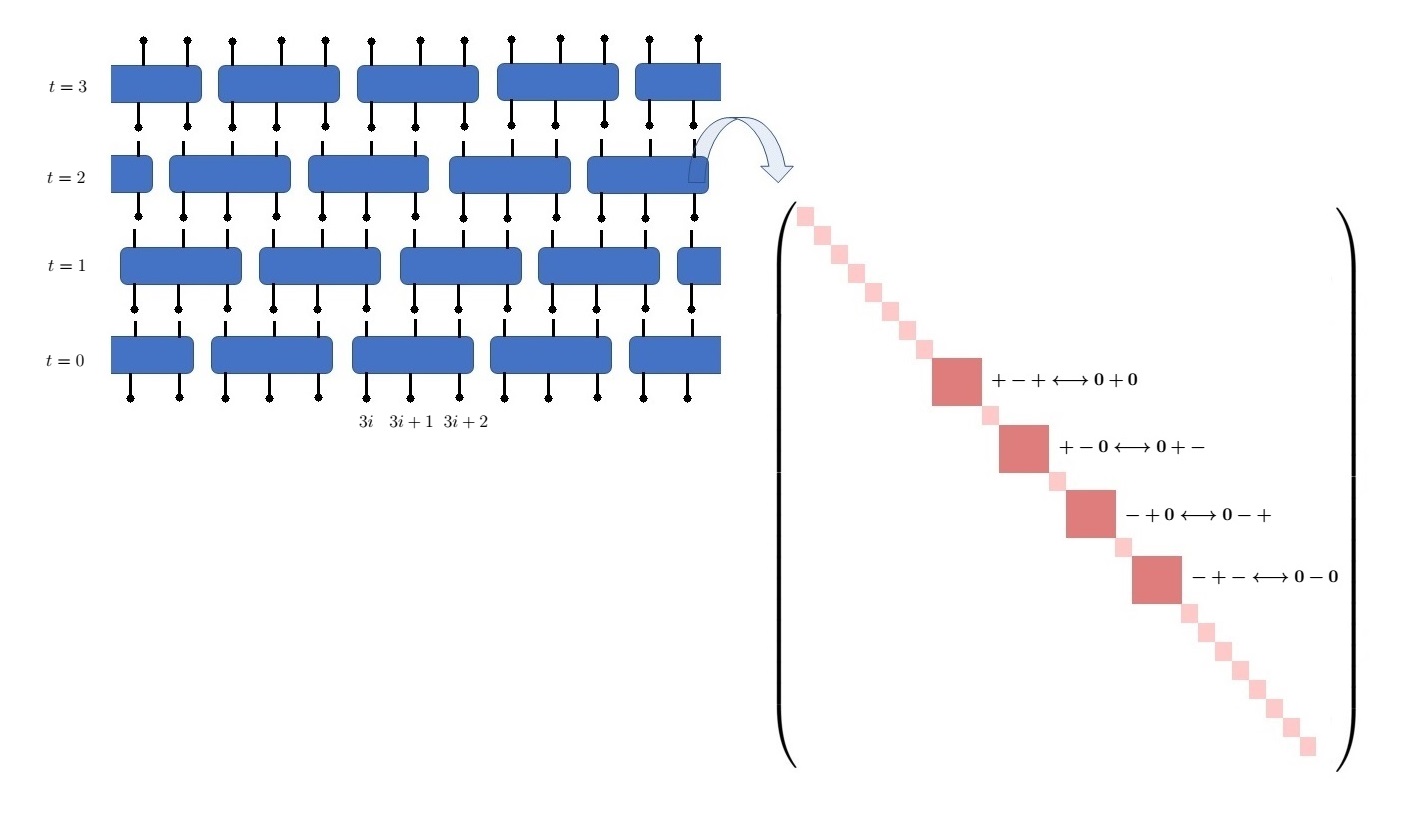}
 \caption{Fractonic random unitary circuit: each site is a three-state qudit. Each gate (blue box) locally conserves charge $Q = \sum_j S^z_j$ and dipole moment $P = \sum_j j S^z_j$ of the three qudits it acts upon. The block diagonal Haar-random unitary with its nontrivial blocks is also shown. Figure taken from \cite{pai2018localization}.}
 \label{fig:floquetCircuit}
 \end{figure}

The simplest realization of these rules is provided by circuits with three site unitary gates, which take the form of $27\times27$ matrices as shown in Fig.\ref{fig:floquetCircuit}. The charge and dipole moment conservation lead to a block diagonal structure in the gates. Notably, there are only four non-trivial two by two `blocks,' each of which is a random unitary drawn independently from the Haar measure on $U(2)$, while the rest of the matrix is diagonal (pure $U(1)$ phase). We will begin our analysis with a discussion of this simple circuit with three site gates, but we will prove that the key results are robust for any finite gate size (while also flagging some special features that are unique to gates of range three). 
We note that while \cite{pai2018localization} considered a circuit that was random in both space and time, this is not important for our purposes - our results hold just as well if the circuit is uniform in space (translation invariant), and/or if it is periodically repeated in time (Floquet). In all that follows, we work with a circuit that is translation invariant, since this makes our central result of localization yet more dramatic. We also work with a circuit that is stroboscopically repeated in time, since this allows us to meaningfully discuss eigenstates. However, we emphasize that our basic results require neither translation invariance in space, nor periodicity in time. 

 This circuit has only two symmetries: charge conservation and dipole moment, and the `symmetry sectors' of the theory are correspondingly labelled by just two quantum numbers: charge $Q$ and dipole $P$. In the Floquet version of the model, three staggered ``layers'' of the circuit are chosen independently, but the layers are then repeated in time. The time evolution operator for one Floquet period is given by 
$U^F  = U_3 U_2 U_1$, where 
\begin{equation}
  U_n=\begin{cases}
    \prod_{i}U^n_{3i,3i+1,3i+2} & \text{if $n = 0$}\\
    \prod_{i}U^n_{3i-1,3i,3i+1} & \text{if $n = 1$}\\
    \prod_{i}U^n_{3i-2,3i-1,3i} & \text{if $n= 2$},
  \end{cases}
\end{equation}
where the gates $U^{1}, U^{2}$ and $U^{3}$ are chosen at random for a given realization, but remain fixed throughout the run corresponding to that realization.  We work throughout with open boundary conditions. In certain layers of the circuit, there may be sites near the boundary that are acted on trivially (pure phase) but the Floquet operator as a whole acts non-trivially on every site. 

\begin{figure}
 \centering
\includegraphics[width=\columnwidth]{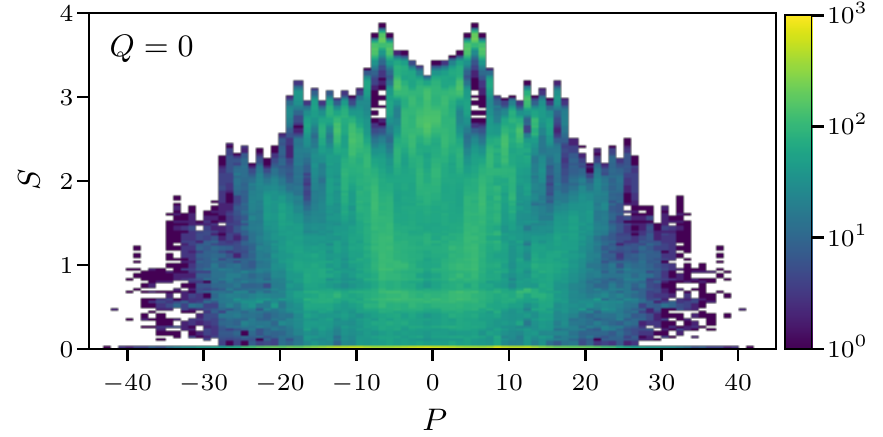}
\caption{ Entanglement entropy of eigenstates as a function of dipole moment for a system with $L=13$ sites, in the symmetry sector with total charge $Q=0$. The color-bar denotes the number of eigenstates with entanglement entropy $S$ and a given $(Q,P)$. For each symmetry sector $(Q,P)$, there is a co-existence of low and high entanglement eigenstates, in sharp contrast to the usual behavior expected from the eigenstate thermalization hypothesis.\label{EE} }
 \end{figure}

Before presenting our analytic proofs,  we  illustrate the unusual properties of this model by numerically studying the eigenstates of the Floquet unitary within each symmetry sector. For each eigenstate $|\psi \rangle$ we construct a density matrix $\rho = |\psi\rangle \langle \psi|$, and extract the half-chain entanglement entropy $S$ according to $S = - \mathrm{Tr}_B \rho \log \rho$, where the trace is over half the chain. In Fig.~\ref{EE} we plot the entanglement entropy of the eigenstates for a system of size $L=13$, in total charge $Q=0$ sector, as a function of dipole moment $P$. We note that the states with maximal charge have $Q=\pm L$, so $Q=0$ corresponds to the middle of the many body spectrum, where we could expect the eigenstate thermalization hypothesis (ETH)~\cite{Deutsch,Srednicki, Rigol} to apply in a translation invariant and not conventionally integrable model. However, in every symmetry sector $(Q,P)$ we find a combination of low and high entanglement eigenstates, in sharp contrast to the usual expectations from eigenstate thermalization, but analogous to the phenomenon of quantum many body scars. As we will show, this apparent violation of the ETH arises from the shattering of Hilbert space. 
 
\section{Shattering of Hilbert space}
\label{sec: analyze}
We now demonstrate how the local constraints fracture Hilbert space, giving rise to an exponentially large number of emergent \emph{dynamical} subsectors that do not mix under the dynamics. By contrast, note that the twin conservation laws of charge and dipole moment only lead to $O(L^3)$ explicit \emph{symmetry} sectors, labeled by the values of charge and dipole moment ranging from $Q = \{-L,  \cdots, L\}$ and $P = \{-\frac{L(L-1)}{2}, \cdots \frac{L(L-1)}{2}\}. $

\subsection{Localized eigenstates}
\label{inert}
In this section, we show how all local fractonic circuits have exponentially many \emph{exactly} localized inert states, labeled by state dependent local integrals of motion (despite the absence of spatial randomness). These constitute emergent subsectors of dimension exactly one. Notably, these inert states are product states of charge ($i.e.$ product states of $S^z$), so these are exceptionally simple, physically realizable states. These states are eigenstates of the Floquet fractonic circuit with zero entanglement, while they are left invariant by circuits that are random (\emph{i.e.} non-repeating) in time, thereby also demonstrating robustness to temporal noise. 

We start with an analytic proof which shows that the combination of $Q,P$ symmetries together with locality is enough to give exponentially many strictly inert states. The construction in our proof is extremely physical, and furnishes a strict lower bound on the number of inert states. Section~\ref{sec: nonfractonic} provides an inductive, though less physical, method which allows us to count the actual number of inert states. 

 \begin{figure}
 \centering
\includegraphics[width=\columnwidth]{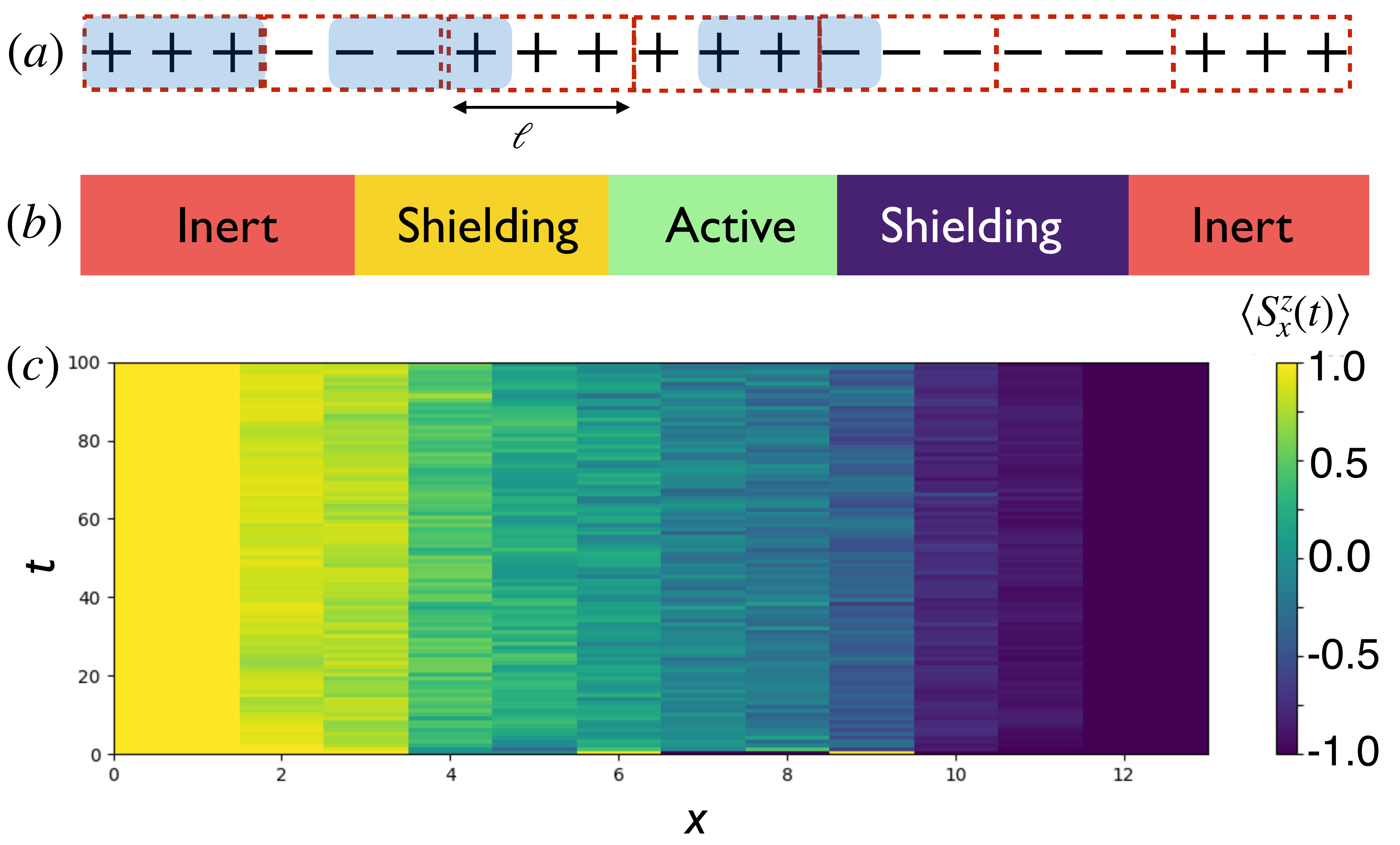}
\caption{ \label{fig:inert} (a) Exponentially many strictly inert states in a model with range $\ell$ can be constructing by dividing the system into size $\ell$ blocks, and randomly picking each block to be of extremal positive or negative charge. A range $\ell$ gate (blue rectangles) acting on such a state locally sees either a configuration of maximal charge, or a configuration of maximal dipole moment for a given charge --- and hence is forbidden from making any local rearrangements. (b) Dynamical subspaces of varying sizes can be constructed by embedding ``active", \emph{i.e.} non-inert, blocks into inert backgrounds. As long as the active block has a finite size, it can be prevented from melting the inert regions by surrounding it with ``shielding" regions of equal or greater size. (c) Dynamics of charge $\langle S^z_x(t)\rangle$ starting from an initial state with a central active region surrounded by shielding regions. We see that the central region thermalizes, but isn't able to melt the boundary spins which remain inert.  }
 \end{figure}

Consider a model with charge and dipole symmetries, and finite range (gate-size) $\ell$. Let us denote by $+$/$-$ the maximum/minimum local charges on a site respectively; these could be the `top' and `bottom' states of a qudit of spin $S$ so that $S^z = \pm S$, or else the occupied and unoccupied states of a hardcore boson model. 
Now, note that {\it any} pattern that alternates between locally `all plus' and locally `all minus,' with domain walls between `all plus' and `all minus' regions at least $\ell$ sites apart, must be inert. These are states of the form $|++++-----++++ \cdots\rangle$ (\emph{cf.} Fig~\ref{fig:inert} (a)). This follows because every gate acting on such a state straddles either zero or one domain walls. If it straddles zero domain walls, then it acts locally on a block with extremal charge, which is obviously inert. If it straddles one domain wall, then it acts on a block with extremal dipole moment {\it given its charge}, and this must also be inert. The inertness of the latter kind of block follows because it is made up of only $+$ and $-$ charged sites, and the only charge conserving moves that one can make are (i) to reshuffle $+$ and $-$ charges and (ii) to lower the charge of a "+" site by 1, and simultaneously raise the charge of a "-" site by 1.  However, if {\it every} $+$ charge is to the right of {\it any} $-$ charge (or vice versa) then any such move necessarily changes the dipole moment, and so is forbidden. 

One can then straightforwardly lower bound the size of the exactly localized subspace for circuits with gate-size $\ell$ by dividing the system up into blocks of length $\ell$, and allowing each block to be either `all plus' or `all minus.' This yields an inert subspace of dimension at least $2^{L/\ell} = c^L$, where $c = 2^{1/\ell}$. This is exponentially large in system size for {\it any} finite gate size $\ell$, and cleanly illustrates how simultaneously conserving charge and dipole moment provably leads to the emergence of exponentially large localized subspaces into which information may be robustly encoded.  

Note that the bound above is not tight; for $\ell=3$ it predicts a localized subspace of dimension at least $1.25^L$, whereas a more careful counting, done in Section~\ref{sec: nonfractonic}, gives a localized subspace of dimension $2.2^L$. Nevertheless, it is sufficient to establish the existence of an exponentially large, robust, localized subspace for any finite gate size. Each of the inert states in this subspace can be labeled by \emph{state-dependent} local integrals of motion corresponding to the local values of charge and dipole moment. Also, note that this type of localization does not require disorder - indeed it occurs even in a circuit that is translationally invariant in the thermodynamic limit and survives temporal noise, as long as the constraints are obeyed.

\begin{figure*}
\centering
\includegraphics[width=\textwidth]{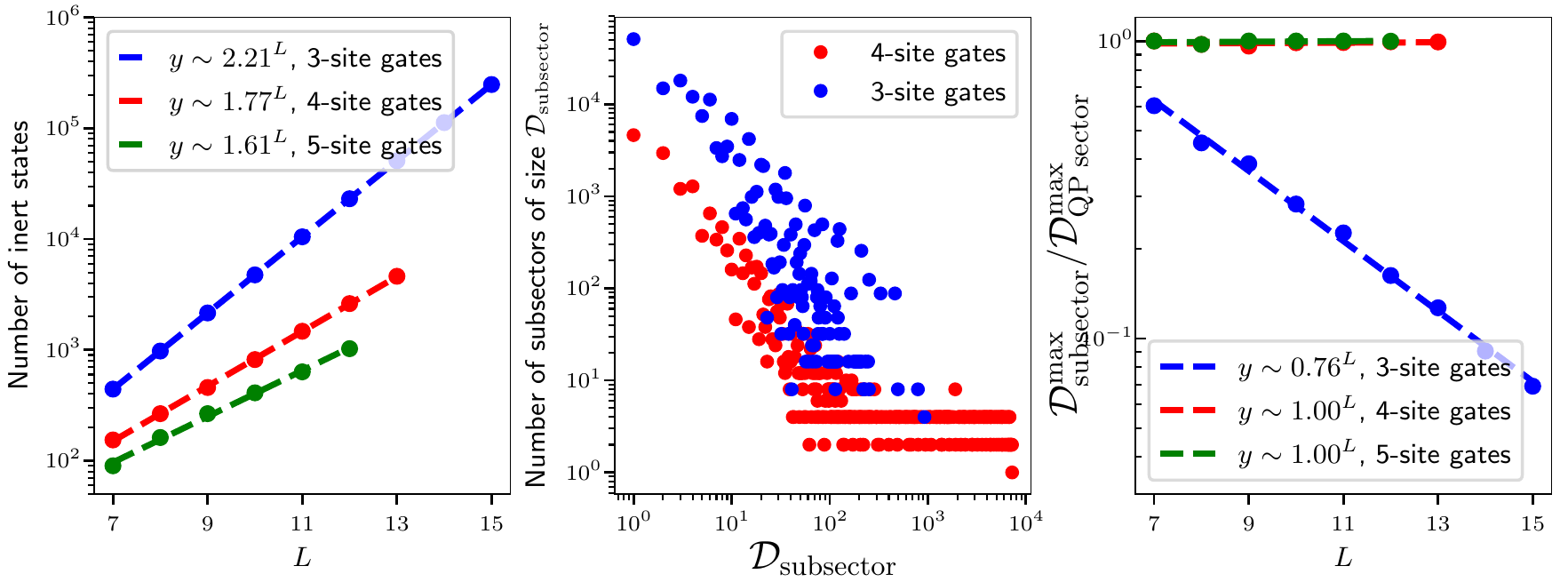}
\caption{(a) Scaling of dimension of inert subspace as a function of system size, for $\ell$ site gates $\ell=3,4,5$. For $\ell=3,4$ the results are consistent with the analytic predictions. For $\ell=5$ we have not worked out an analytic prediction, but the results are consistent with the lower bound established in Sec.~\ref{inert}.  (b) Plot showing subsector size distribution. For three site gates the frequency of subsectors of a particular size decreases polynomially with the Hilbert space dimension of the subsector. For four site gates there is an initial polynomial decrease followed by a saturation, in that beyond a certain subsector size, further increases in subsector size do not seem to translate into a decrease in frequency. Notably, the maximum size of the emergent subsectors for four site gates is much larger than that for three site gates. (c) Plot showing the relative sizes of the largest emergent subsector and largest $(Q,P)$ symmetry sector, which corresponds to $Q=0, P=0$.   For three site gates, the size of the largest emergent subsector is a vanishing fraction of the size of the largest symmetry sector, in the thermodynamic limit, whereas these sizes scale similarly for four and five site gates, showing that the fracturing is more severe for three site gates.  \label{longrangegates}}
\end{figure*}

\subsection{Larger subsectors}
\label{larger}
We now turn to a systematic construction of emergent dynamical subspaces of dimension greater than one, which do not mix with the rest of the Hilbert space. The main idea is to build subspaces of various sizes by embedding ``active" (non-inert) blocks into inert backgrounds, and appropriately ``shielding" the active blocks so as to keep the active region localized in a finite region of space (Fig.~\ref{fig:inert}(b)). The size of the sector so built will be controlled by the Hilbert space dimension of the active blocks, and 
we can embed multiple active blocks spatially separated by inert regions. Strikingly, this leads to a coexistence of spatial regions that thermalize or not, starting from a single initial state! This is different even from the case of scars in other models like the PXP where the thermalization, or lack thereof, is controlled by the initial state but there is no further spatial dependence of the relaxation of observables.  

To illustrate, Fig.~\ref{fig:inert}(c) shows the expectation value of the charge $\langle S^z_x(t)\rangle$ in a system of length $L=14$, initialized in a state with a central active region surrounded by shielding regions (explained below). We can see that although the spins at the center thermalize, they never succeed in entirely melting the shielding region, so that the spins on the boundary of the system remain frozen throughout the time evolution! In other words, the shielding regions can protect the boundary spins against decoherence, despite the presence of the fluctuating active region nearby. In this example, the inert spin lies at the boundary merely for ease of depiction in a finite size system --- this chunk of 14 sites can be embedded into a larger system by extending the inert configurations on either end. 

At this point, one may wonder if the `embedding' of active regions into inert subspaces actually works for `active' regions of arbitrary size, or if there is a critical size of active region beyond which the problem `avalanches' \cite{avalanche}, causing the entire inert region to `melt.' However, it is straightforward to prove that {\it any} finite size of active region can always be contained by suitably chosen finite sized shielding regions. For example, take any finite sized active region, and flank it with `shielding' regions that are `all plus' to the right, and `all minus' to the left, and which are at least as large as the active region (cf. Fig.~\ref{fig:inert}). Now the active region can start to `melt' the shielding regions, but in doing so it will inevitably either be moving plus charge left, or minus charge right, both of which reduce the dipole moment. To preserve dipole moment overall, the active region would have to increase its internal dipole moment to compensate. However, a {\it finite} sized active region has a maximum internal dipole moment that it can accommodate, and as such the active region {\it cannot} entirely melt suitably chosen `shielding' regions of the same size. Outside the shielding regions, the state can then remain inert, as in Fig.~\ref{fig:inert}(c). At a technical level, the problem avoids avalanches \cite{avalanche} because as the `active' region grows, it has to increase its dipole moment and become less active. Consequently, one may embed active regions of any desired size into the inert subspace, by choosing the appropriate shielding. 

\begin{figure}
\centering
\includegraphics[width=\columnwidth]{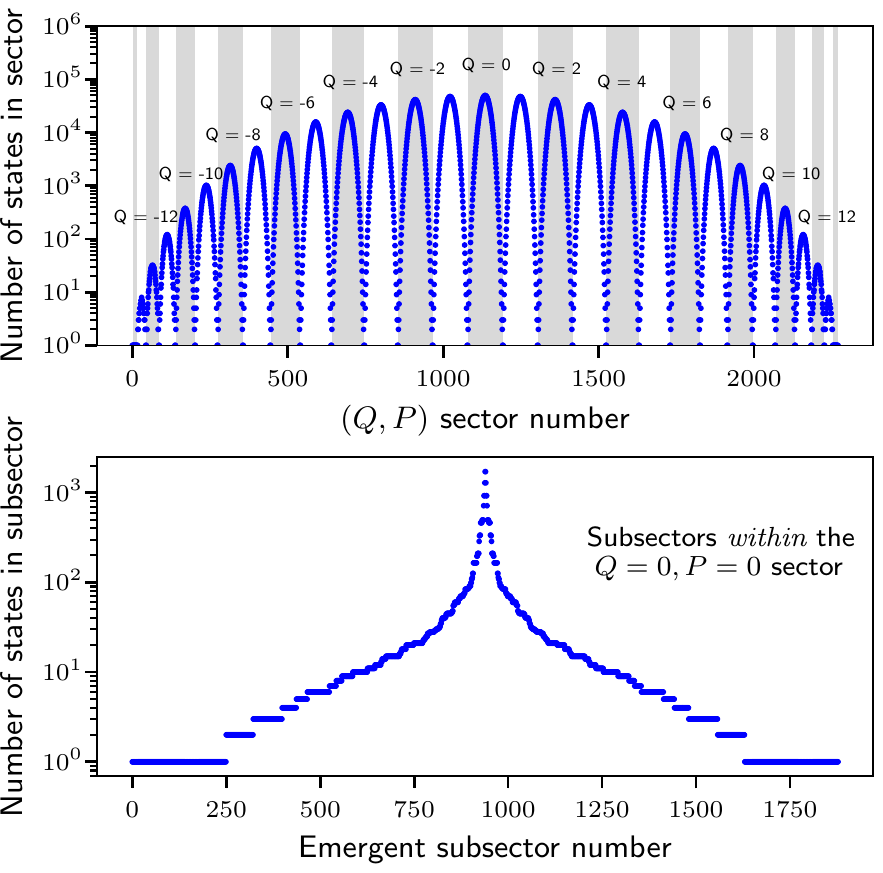}
\caption{\label{fig:subsectors} (a) Breakup of the Hilbert space into symmetry sectors labelled by charge $Q$ and dipole moment $P$ (b) Further shattering of each symmetry sector into emergent subsectors of various size, here shown for the symmetry sector with $(Q,P) = (0,0)$. }
\end{figure}

We have therefore proven that the Hilbert space within each symmetry sector `shatters' into numerous emergent subsectors of all sizes. This `shattering' may be straightforwardly verified numerically extracting the `connectivity' of the Floquet operator, within a particular symmetry sector. In Fig.~\ref{fig:subsectors} we show this shattering quantitatively, for a twelve site system in the sector with $Q=0$ and $P=0$ and three-site gates. The sectors with exactly one state correspond to the `inert' states (localized subspace) discussed above, but as one can see, there is a distribution of emergent subspaces of a wide variety of sizes. In Figure~\ref{longrangegates}(b), we show the full distribution of emergent subsector sizes for circuits with gate size $\ell = 3,4$ in a system of size $L=13$ with spin 1 degrees of freedom on each site. The figure shows that the frequency of subsectors of a particular size decreases polynomially with the dimension of the subsector, followed by a saturation. 

The broad distribution of emergent subsector sizes largely explains the broad distribution of eigenstate entanglement entropies found within a given $(Q,P)$ symmetry sector. Indeed, the eigenstate entanglement for a given cut is controlled by size of the dynamical subsector in which the eigenstate lives (more specifically, the size of the largest active block straddling the entanglement cut), and \emph{not} the dimension of the full symmetry sector. This is discussed further in Appendix~\ref{app:entanglement}.

We now turn to an important distinction between three site gates vs. gates of size four and larger. The largest subsector for three-site gates is numerically observed to contain exactly $\binom{(L-1)}{\floor{(L-1)/2}}$ states, which asymptotically scales as $2^{L}$. This is a vanishing fraction of the largest symmetry sector labeled by a particular quantum number for $(Q,P)$, which scales as $3^L$ (upto polynomial in $L$ corrections). This indicates a strongly constrained dynamics, which is only ever able to connect a vanishing fraction of the full Hilbert space, also shown quantitatively in Figure~\ref{longrangegates}(c). This scenario is referred to as `strong' fracture of the Hilbert space. 
By contrast, the figure shows that the largest dynamical subsector with longer range gates asymptotically has the same size as the largest symmetry sector, denoting `weak fracture', and thus the dynamics can access much larger parts of the Hilbert space. 

Intuitively, the distinction between three and four site gates can be understood due to the presence of ``bottlenecks" in the  range three system, which refer to finite motifs that `cut' the chain in two, regardless of the state these motifs are embedded into -- so that the two halves of the system on either side of the bottleneck become dynamically disconnected (see Appendix~\ref{bottlenecks} for details). In contrast, systems with range four and larger do not have such bottlenecks. The reason is that if one has a large sea of zeros `00000', then `vacuum fluctuations' of this can pair produce anti-aligned dipoles  `-++-', which can then separate and travel freely and destroy bottlenecks. While such fluctuations are not possible with range three gates and spin 1, one can get rid of bottlenecks upon considering larger spins, say $S=2$.

The distinction between strong and weak fracture has important consequences for dynamics. While exponentially many strictly inert (or mostly localized) pure states exist for gates of any finite range, dynamics from a randomly chosen \emph{typical} initial state is expected to be highly sensitive to the degree of shattering. If the largest dynamical subsector is a vanishing fraction of the full Hilbert space, as in the case of three site gates, then dynamics from a randomly chosen initial state will be non-ergodic at all times. This is in contrast to weakly fractured systems where typical initial states have some weight in the largest dynamical subsector and this contains most of Hilbert space, such that the dynamics from randomly chosen initial states can also explore most of Hilbert space. These distinctions are discussed further within the context of entanglement dynamics in App.~\ref{app:dynamics}. This discussion also explains the numerical results of Ref.~\cite{pai2018localization} which observed localization in operator dynamics (which averages over \emph{all} states) for three site gates, but not four site gates~\cite{erratum}.

More detailed implications of these shattering and shielding phenomena are discussed in the Appendices. 

\section{Higher dimensions}
\label{sec: HigherD}
Thus far we have restricted our discussion to systems in one space dimension. We now discuss how the results may be extended to higher dimensional systems on hypercubic lattices. 

\subsection{Shattering from dipole conservation}
\begin{figure*}
    \includegraphics[width=\textwidth]{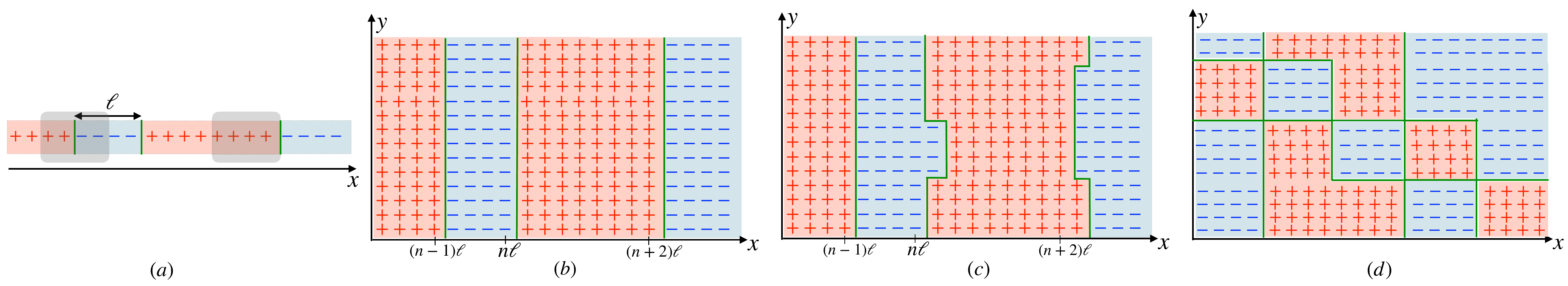}
    \caption{Inert states in one and two dimensions. Thick green lines indicate domain walls. (a) The state is divided into blocks of length $\ell$, and each block is chosen randomly to be $+$ or $-$. Each such pattern is an eigenstate of the dynamics when $Q$ and $P_x$ are conserved because a range $\ell$ gate (gray blocks) either sees a pattern of extremal local charge or extremal local dipole moment.  (b) Translating one dimensional inert states along $\hat y$ still gives inert states (c) When $Q$, $P_x$ and $P_y$ are conserved in 2d, the domain walls can be allowed to `roughen', while the state still remains an eigenstate of the dynamics. (d) When charge, dipole and quadrupole are conserved, inert states are obtained by dividing the system into blocks of size $\ell \times \ell$, and picking each block to be $+$ or $-$.}
    \label{fig: inert}
\end{figure*}

We begin with the case where the dynamics is generated by gates that conserve charge and also conserve all components of dipole moment, defined along the various lattice directions as $P_\alpha = \sum_{\mathbf{r}} r_\alpha S_\mathbf{r}^z$, where $\mathbf{r} = (r_x, r_y, r_z \cdots)$. In one dimension, we have proven that the conservation of $Q$ and $P_x$ provably gives exponentially many (at least $2^{\frac{L}{\ell}}$) strictly localized `inert' states that are left invariant by the dynamics (Sec.\ref{inert}). These are obtained by considering product states which are either $+$ or $-$ on {\it every} lattice site, with $+$ and $-$ regions separated by domain walls that 
are at least $\ell$ sites apart (Fig.~\ref{fig: inert}(a)).  Crucially, even though such states can have the same \emph{global} $Q, P_x$ quantum numbers as other states, they cannot be connected to these other states under \emph{local} dynamics. 

We now turn to higher dimensions $d>1$. Of course, the 1d states considered above can be extended in a translationally invariant fashion in directions orthogonal to $\hat x$ (Fig.~\ref{fig: inert}(b)) and all such states would still be inert. But there are only exponentially many in $L$ such states. However, we now show that if \emph{all} components $\{P_\alpha\}$ are conserved, then the number of inert states is $\sim \exp(c L^d)$. For specificity, consider a system in $d=2$ space dimensions ($\hat x$, $\hat y$). Start with a `stripe' state with domain walls parallel to the $y$ axis and at least $\ell$ sites apart in the $x$ direction. 
Now, note that these domain walls can be allowed to `roughen' slightly while leaving the state inert. For specificity: divide up a domain wall that lives between sites with $x$-coordinate $n\ell$ and $(n\ell+1)$ into blocks of length $\ell$. In each such block, allow the domain wall to uniformly shift in the $+\hat x$ direction by either zero or one lattice spacings (Fig.~\ref{fig: inert}(c)).  
This reduces the spacing between domain walls by at most one in the $\hat x$ direction; to wit, all sites with $n\ell+1< x \leq (n+1)\ell$ are $+$ while all sites with $(n-1)\ell+1< x \leq n \ell$ are $-$, and dipole conservation of $P_x$ still prohibits any rearrangement involving these sites. It is only along the line $x=(n\ell+1)$ that we encounter both $+$ and $-$ sites, and can make rearrangements that conserve the $\hat x$ component of dipole moment. However, along this line we see alternating $+$ and $-$ regions with domain walls at least $\ell$ apart, and conservation of $P_y$ guarantees that this too must be inert. Thus, in fact any such `roughened' configuration of a domain wall is inert. There are $N(L) \sim 2^{L/\ell}$ inert `roughened' configurations of each domain wall, and $L/\ell$ places where we could choose to place a domain wall (or not), so the total number of inert states is at least $\sim N(L)^{L/\ell} \sim 2^{L^2/\ell^2}$. This argument proceeded by a dimensional reduction to the one-dimensional problem. By the same token, the argument extends to hypercubic lattices in arbitrary dimension, so that conservation of charge and all components of dipole moment is always sufficient to guarantee the existence of an exponential in volume number of inert states.

\subsection{ Shattering from multipole conservation}
 The localized states considered above were fundamentally one dimensional (i.e. `stripe-like'). We now consider a class of {\it intrinsically} d-dimensional localized states, with number exponential in system volume, which become available if the system conserves the first $d$ multipoles of charge. 

Consider a two dimensional system on a square lattice constrained to conserve charge $q$, dipole moment $\{P_\alpha\}$, and also {\it quadrupole moment} $\{P_{\alpha \beta}\}$. Note that in two dimensions quadrupole moment is a rank two symmetric traceless tensor with two independent entries $P_{xy}$ and $P_{xx} - P_{yy}$, corresponding to the `dipole of a dipole' in the directions perpendicular and parallel to the dipole vector respectively. The components of $P_{\alpha \beta}$ are defined in $d$ space dimensions as $P_{\alpha \beta} = \int dV \rho(r) [d r_\alpha r_\beta - r^2 \delta_{\alpha \beta}]$, where $dV$ indicates a volume integral, $\rho$ is the charge density, $\delta_{ij}$ is the Kronecker delta function, and the definition depends on the choice of origin (with obvious lattice generalizations)  Assume that charge, dipole, and $P_{xy}$ are all locally conserved (conservation of $P_{xx}-P_{yy}$ is not necessary). Now consider `checkerboard' states made by dividing up the system into $\ell \times \ell$ squares, and allowing every square to be either all $+$ or all $-$ randomly (Fig.~\ref{fig: inert}(d)). There are $2^{L^2/\ell^2}$ such states. Any gate acting on such a state acts across either zero or one corners. If it acts across zero corners, then it acts on a state which is locally either maximum charge, or maximum dipole given its charge, and charge and dipole conservation suffices to guarantee that the gate must act trivially (i.e. as a pure phase). Meanwhile, if the gate acts across one corner, then it acts on a state that is locally of extremal $P_{xy}$ given its charge and dipole moment, such that charge, dipole and $P_{xy}$ conservation again forces the gate to act trivially. It then follows that every `checkerboard' state of this form is an exact eigenstate of the dynamics, concluding our proof that there is an exactly localized subspace of dimension at least $2^{L^2/\ell^2}$. Likewise, one can generalize to three and higher dimensions, so long all higher multipoles of charge are conserved. For instance, three dimensions would require conservation of charge, dipole moment, all off diagonal components of the quadrupole moment ($P_{xy}, P_{xz}, P_{yz}$), and the fully off diagonal ($P_{xyz}$) component of the octupole moment -- in which case all $2^{L^3/\ell^3}$ cubic tilings of space with maximal local charge would be inert.

\subsubsection{Shattering and shielding with multipolar conservation laws} We now show that in addition to the `exactly localized' subspace discussed above, there also arises a broad distribution of dynamical subsectors of various sizes, similar to the 1d case. Upon embedding active regions in inert states, one  can prevent an `avalanche' by simply surrounding the active region by a `shielding' region of maximal multipole.  In two dimensions, a suitable shielding region would be all $+$ in the first and third quadrant, and all $-$ in the second and fourth quadrant. As before, any process by which the active region `melts' the shielding region necessitates increasing the multipole moment of the active region (dipole moment in one dimension, $P_{xy}$ in two dimensions, etc), which makes the active region less active. Moreover, the deeper the rearrangements extend into the shielding region, the {\it bigger} the change in the multipole moment in the shielding region, and thus the bigger the back action on the active region. For a finite sized active region and a sufficiently large finite shielding region, the time evolution operator  cannot have any matrix elements to product states which differ from the initial condition at the outer boundaries of the shielding region, or beyond. The initial condition can thus only mix with a finite number of other product states. In this manner, one may construct dynamical subspaces of a range of sizes by embedding one or more finite volume active regions into the otherwise localized subspace, hence `shattering' Hilbert space. 

\section{Shattering in non-fractonic circuits}
\label{sec: nonfractonic}

Thus far, our discussion of circuits exhibiting shattering has been particular to circuits with `fractonic' constraints (viz. conservation of charge and certain multipole moments thereof).  However, not obviously fractonic circuits displaying a similar shattering of Hilbert space may also be constructed. To present such examples, it is instructive to first consider a more precise counting of the inert states in fractonic circuits using an inductive method. This leads to a natural generalization to non-fractonic examples. 

We start with circuit with range three unitary gates. For system size $L=3$, there is only one gate acting, and there are exactly $19$ product states (in the charge basis) which have trivial dynamics, and are hence localized - these are the $19$ states acted upon by trivial blocks of the constrained random unitary in Fig.\ref{fig:floquetCircuit} (e.g. the state $|00+\rangle$). These states do not mix with the rest of the Hilbert space, and are hence `inert,' lying in a subsector with dimension one. Meanwhile, if a state is inert  in a system of size $L$, then it will remain inert when an additional degree of freedom is added if the final two degrees of freedom of the $L$ site system and the additional degree of freedom collectively form one of the `inert' configurations of an $L=3$ site system. This is because the only ``new" dynamics in the presence of the additional spin comes from the addition of a single three site unitary gate acting on the three spins formed by the added spin and the two penultimate spins of the length $L$ chain. Importantly, for {\it any} inert state of an $L$ site system, there is at least one choice of spin state for the added spin (and sometimes more than one), which leaves the resulting state in the $L+1$ site system also inert. Specifically, an inert state in a system of size $L$ remains inert upon addition of another degree of freedom if the conditions tabulated in Table \ref{threesiteconstraints} are satisfied. Now let $N_{ab}(L)$ be the number of inert states in a system of size $L$, in which the final two sites have $S^z$ eigenvalues $a$ and $b$ respectively. The total number of inert states for a system of size $L$ is obtained by summing $N_{ab}(L)$ over all choices $ab$. Using Table \ref{threesiteconstraints} , we can see that these quantities obey the recursion relations

\begin{eqnarray}
\left( \begin{array}{c} N_{++}\\ N_{+0}\\ N_{+-}\\ N_{0+}\\N_{00}\\N_{0-} \\N_{-+} \\N_{-0}\\N_{--}
\end{array}
\right)_{L+1} = \left(\begin{array}{ccccccccc} 1 & 0 & 0& 1 & 0 & 0 & 1 & 0 & 0 \\ 1 & 0 & 0 & 0 & 0 & 0 & 0 & 0 & 0 \\ 1 & 0 & 0 & 0 & 0 & 0 & 0 & 0 & 0 \\ 0 & 1 & 0 & 0 & 1 & 0 & 0 & 1 & 0 \\ 0 & 1 & 0 & 0 & 1 & 0 & 0 & 1 & 0 \\ 0 & 1 & 0 & 0 & 1 & 0 & 0 & 1 & 0\\ 0 & 0 & 0 & 0 & 0 & 0 & 0 & 0 & 1\\ 0 & 0 & 0 & 0 & 0 & 0 & 0 & 0 & 1\\ 0 & 0 & 1 & 0 & 0 & 1 & 0 & 0 & 1\end{array} \right)
 \left( \begin{array}{c} N_{++}\\ N_{+0}\\ N_{+-}\\ N_{0+}\\N_{00}\\N_{0-} \\N_{-+} \\N_{-0}\\N_{--}
\end{array}
\right)_{L} \nonumber
\end{eqnarray}
This matrix can be diagonalized and its eigenvalues and eigenvectors, combined with the known values for $N_{ab}(3)$ can be used to exactly determine the number of inert states for any $L$. However, asymptotically at large $L$, the growth will be controlled by the largest eigenvalue of this matrix, $\lambda$, i.e. the dimension of the Hilbert space grows asymptotically as $|\lambda|^{L}$. The matrix in question has only one real, positive eigenvalue with norm greater than one, $\lambda \approx \ 2.2$ which tells us that the dimension of the localized subspace grows asymptotically as $\sim 2.2^L$. 
\begin{table}
\begin{tabular}{c|c}
\underline{Last two sites of L site chain are} & \underline{Site added can be} \\
$+$ $+$ & $+$ or $0$ or $-$\\
$+$ $0$ & $+$ or $0$ or $-$\\
$+$ $-$ & $-$ \\
$0$ $+$ & $+$\\
$0$ $0$ & $+$ or $0$ or $-$\\
$0$ $-$ & $-$\\
$-$ $+$ & $+$ \\
$-$ $0$ & $+$ or $0$ or $-$\\
$-$ $-$& $+$ or $0$ or $-$
\end{tabular}
\caption{For the fractonic circuit with three site gates, if an inert state in a system of size $L$ has the final two sites in the states shown in the left column, then it remains inert upon addition of another spin if the new spin is in the corresponding state shown in the right column. \label{threesiteconstraints}}
\end{table}

We therefore conclude that in the thermodynamic limit there are  approximately $2.2^L$ inert states, each of which exists in its own emergent subsector, undergoes trivial (pure phase) dynamics, and does not mix with the rest of the Hilbert space. This is verified by exact numerical counting of the number of inert states in systems upto sizes $L=15$, and shown in Fig.~\ref{longrangegates}(a). 

We note that the key feature of fractonic circuits that leads to this exponentially growing inert subspace is the existence of \emph{multiple} pathways or choices for getting new inert states upon adding spins to inert states of a given size. By contrast, in a system with only charge conservation,  the only choices for building inert states require $++$ to be followed by $+$, or $--$ to be followed by $-$. This, however, gives exactly two inert states due to a lack of exponential branching arising from multiple pathways.

Generalizing this, one can verify via a similar asymptotically exact counting (see Appendix~\ref{app:foursitecounting}) that a fractonic circuit with four site gates also has an exponentially large localized subspace, with asymptotic dimension $\sim 1.8^L$ in the thermodynamic limit, again numerically verified in Fig.~\ref{longrangegates}(a). Exact analytical calculations for larger gate sizes rapidly become tedious, but the construction depicted in Fig.~\ref{fig:inert} is sufficient to show that an exponentially large exactly localized subspace survives for any finite gate size $\ell$.

Let us now turn to non-fractonic examples, building on the construction above.  Consider a circuit made out of local two spin gates acting on a one dimensional chain of $S=1$ spins. If this two site gate is constrained so that it acts trivially on the states $|0+\rangle$, $|+0\rangle$, $|0-\rangle$, and $|-0\rangle$, then it may be readily verified, through methods similar to those above, that there is an exponentially  large space of inert states displaying trivially localized dynamics. For a chain of size $L=2$, there are then exactly four inert states. Meanwhile, if $N_{\beta}(L)$ is the number of inert states ending in $\beta$ in a system of size $L$, then this quantity obeys the recursion relation
\begin{eqnarray}
\left( \begin{array}{c} N_+ \\ N_0 \\ N_-
\end{array}
\right)_{L+1} = \left(\begin{array}{ccc} 0 & 1 & 0 \\ 1 & 0 & 1 \\ 0 & 1 & 0 \end{array} \right)
 \left( \begin{array}{c} N_+ \\ N_0 \\ N_-
\end{array}
\right)_{L} \nonumber 
\end{eqnarray}
The matrix in the recursion relation has eigenvalues $\pm \sqrt{2}$ and zero. The dimension of the degenerate subspace thus grows asymptotically as $\sqrt{2}^L$, providing a concrete example of a not obviously fractonic circuit with an exponentially large localized subspace. The mechanism again involves the existence of ``multiple" pathways for extending inert states when new sites are added.  However, in the absence of a physical principle giving rise to this particular circuit architecture, analogous to the `fractonic' constraints of charge and dipole moment conservation, it is unclear how this circuit should be generalized to gates of longer range, and hence the question of whether this `shattering' survives in the presence of longer range gates is ill posed. Nevetheless, `shattering' may be produced by similar constructions in circuits involving gates of larger size - a sufficient condition is that there should exist at least two locally inert patterns which can be combined together in an inert fashion. 

A fruitful perspective on which types of circuits produce `shattering' of Hilbert space is provided by recursion relations of the form discussed above. For a circuit acting on a system with local Hilbert space dimension $q$, and random $N$ site gates, the recursion relation is governed by a square matrix of size $q^{N-1}$. The entries in this matrix can only be $0$ or $1$ - and at least two of the entries must be zeros, otherwise the circuit acts trivially on every possible state (which is a trivial shattering, say by diagonal matrices). {\it Every} such matrix with an eigenvalue larger than $1$ specifies a circuit with an exponentially large inert subspace. From this it follows that there are no spin $1/2$ chains with only two site gates that realize a shattered Hilbert space (in the obvious $z$ basis)- spin $S=1$ and two site gates is the minimal case necessarily to realize such shattering.

\section{Physical Realizations}
\label{sec: physical}
\begin{figure*}
    \includegraphics[width=.9\textwidth]{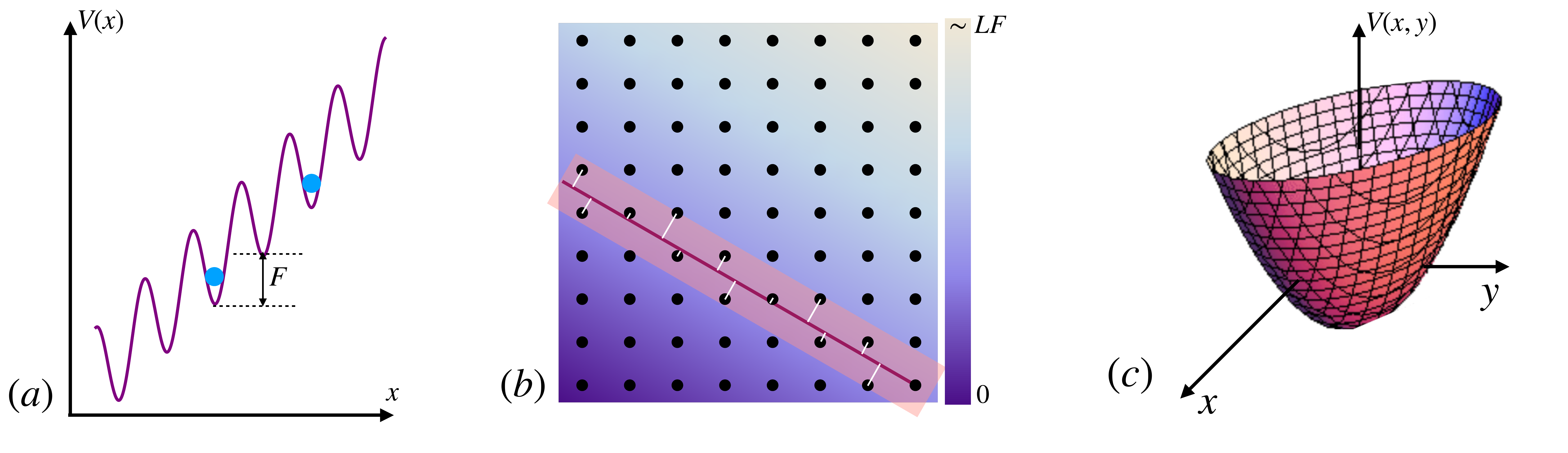}
    \caption{How to obtain dipole and quadrupole conservation. (a) A one dimensional system in a `tilted potential' conserves dipole moment upto an exponentially long prethermal timescale. (b) When a `tilted potential' is applied at an angle in two dimensions, dipole moment in the direction of the tilt is conserved but one has to worry about hopping in an `equipotential shell' perpendicular to the tilt direction (shown). The color scale denotes the potential energy due to a tilt at 60 degrees relative to the $\hat x$ axis. Sites within the equipotential shell see a quasiperiodic potential set by the distance from the equipotential line. (c) Quadrupole conservation in two dimensions may be obtained by placing the system in a harmonic trap, with inequivalent trap frequencies along two orthogonal directions, and with the trap rotated with respect to lattice axes.}
    \label{fig:realize}
\end{figure*}

We now discuss how conservation laws on multipole moments, and the associated localization from shattering, may be generated in physically realistic settings. Dipole moment couples directly to electric field, so if the system is placed in a sufficiently large static electric field (or equivalently, a tilted potential) then the Hilbert space will (approximately) split into symmetry sectors labelled by dipole moment (equivalently, center of mass position), with states at different dipole moment having sharply different energies. A minimal model for realizing approximate conservation of charge and dipole moment in one dimension would thus be given by a model of hardcore bosons (or spinless fermions) in a linearly varying scalar potential (Fig.~\ref{fig:realize}(a)) so that $H = H_0+V$:
\begin{eqnarray}
H_0 &=& J \sum_{\langle x \rangle} b^{\dag}_x b_{x+1} + U \sum_i n_x n_{x+1} \\
V &=& F \sum_x x n_x = F P_x
\end{eqnarray}
where $n_x=1$ ($n_x=0$) corresponds to the $+$ ($-$) state \emph{i.e.} the presence or absence of a boson on site $x$. This model can be mapped to a spin 1/2 system with spin up and down states mapping to $n_x = 1,0$ respectively. Likewise, we could choose $H_0$ to be a standard Fermi-Hubbard Hamiltonian with spinful fermions, in which case the $n_x = \{0,1,2\}$ states can be mapped to the $S^z$ states of a spin-1 qudit, with the extremal $\mp$ states corresponding to empty ($n_x=0$) and doubly occupied ($n_x = 2$) respectively. The only explicit conservation laws in this system are energy and charge. 

Now if one prepares an initial state at high temperature with repect to $H_0$, say a charge density wave (CDW) of small amplitude $A$ and wavenumber $k$ with $\langle n(x, t=0)\rangle = \overline{n} + A \cos(k x+ \phi)$, we (naively) expect the system to evolve towards towards thermal equilibrium by exchanging energy between the tilt and `non-tilt' parts of $H$~\cite{BakrTilted}. However, we note that the spectrum of $F$ is \emph{super}-extensive ($\sim L^2$) while that of $H_0$ is merely extensive. Thus, if we prepare an initial state with $\langle P_x\rangle_0 \sim L^2$, the tilt energy cannot be dissipated through $H_0$ and the final equilibrium state maintains the same value of $\langle P_x\rangle$, regardless of $F$ (modulo $O(L)$ corrections). In contrast, for initial states with merely extensive $\langle P_x\rangle_0\sim L $, the tilt energy \emph{can} potentially be dissipated by heating the system towards infinite temperature with respect to $H_0$ which would take the system towards a uniform density profile at late times (in a possibly subdiffusive manner~\cite{BakrTilted}). 
For such states, we expect the behavior to be qualitatively different for large and small $F$, and the relaxation towards infinite temperature is only appropriate for small $F$. 

Let us now consider large $F$. In this case, the tilt is the dominant term in $H$ and the spectrum splits into sectors labelled by dipole moment which now becomes a conserved quantity. Of course, this is only {\it strictly} true at infinite $F$ (or if $F\sim L$). However, even when $F$ is finite so that sectors with different values of $P_x$ overlap, it is possible to obtain long-lived approximate conservation of $P_x$ provided $F \gg J$. In this case, rearrangements of the system that change $P_x$ can only occur at a high order $\sim (F/J)$ and one can appeal to the theory of prethermalization~\cite{Rosch, abaninpretherm, MoriPrethermal1} which predicts that $P_x$ will be conserved up to an exponentially long timescale $t_* \sim \exp(F/J)$~\footnote{This theory requires the spectrum of the conserved operator, in this case $P$, to have a discrete harmonic spacing which is true of $P$}. 

On timescales short compared to $t_*$, the system is described by an `effective Hamiltonian' that is constructed as a power-series in $1/F$, and which displays an emergent dipole conservation for an operator $\tilde{P_x}$ which agrees with $P_x$ to leading order. 
The terms in the effective Hamiltonian which are off diagonal in the local density basis must conserve center of mass. The lowest order off diagonal term is of the form $b^{\dag}_i b_{i+1} b_{i+2} b^{\dag}_{i+3} + h.c.$. If we stop here then the effective Hamiltonian exhibits `strong fracture' in that a finite size motif (e.g. a string of five consecutive occupied sites) can `cut' the chain in two, with no charge transport being possible across this motif.  In this case a {\it typical} state will exhibit localized charge dynamics, with charge being exponentially unlikely to wander far from its initial position. Moreover, a density wave that alternates between $+$ and $-$ with domain walls at least four sites apart will be an {\it exact} inert eigenstate of the dynamics. This is however an approximation to the dynamics obtained by truncating the expansion at the lowest non-trivial order in $1/F$. 

At higher orders, the effective Hamiltonian acquires terms of all possible spatial ranges, with terms of spatial range $R$ (assumed large) being generated at order $a R$ in perturbation theory, where $a$ is an $O(1)$ number. Such terms will then have amplitude $(J/F)^{a R}$, and will become important on a timescale $t(R) \sim \exp(a R \log(F/J))$. Once the longer range terms are incorporated into the Hamiltonian, the fracture will be only `weak' (i.e. the size of the largest dynamical subsector will scale the same way as the size of the symmetry sector, upto pre-exponential corrections) and so a {\it typical} state will only exhibit localized dynamics up-to a timescale set by the least weak off diagonal term of range larger than some critical value $R_c \geq 4$. This timescale will be $t(R_c)$. However, an initial condition that alternates between $n_i=1$ and $n_i = 0$ with domain walls at least $\ell$ sites apart will remain an exact eigenstate of the dynamics up to the timescale $t_c \sim \min( t(\ell)), t_*)$, beyond which longer range terms or dipole non-conserving effects will become important. On timescales longer than $t_*$ we expect the system to thermalize to the energy density set by the initial condition. 
Finally, if $t(\ell) < t_*$ then on the intermediate times-cales the thermalization will be to an energy shell {\it and restricted to a certain dipole sector}, and this may have interesting features that are beyond the scope of the present work. 

To summarize, the sharpest signature of fracturing at large tilts is a strong initial state dependence in the dynamics. States that would be strictly inert with exact dipole conservation (such as a CDW with \emph{maximal} amplitude for density fluctations with wavelengths greater than $\ell$) will still look inert, albeit only upto a long time-scale $t_c$. On the other hand, a state with small overlap on the inert states (say CDW states with small amplitudes for density fluctuations) will relax towards a uniform density profile, perhaps subdiffusively~\cite{BakrTilted}. We note that $t_c$ may look infinite in a finite-sized system for which a large enough tilt could lead to actual --- rather than prethermal --- conservation of $P_x$.  

This initial state dependence also emphasizes that the origin of the observed Bloch/Stark MBL at large tilts is entirely distinct from the usual MBL phenomenology in disordered systems, which relies on the existence of exponentially many emergent integrals of motion and predicts localization for any typical initial state. Instead, the numerical observations of Stark MBL~\cite{pollmannstark, refaelbloch} follow from Hilbert space shattering. To wit, the main diagnostics presented in ~\cite{pollmannstark, refaelbloch} were (i) a lack of level repulsion in the energy spectrum which is explained by the presence of exponentially many emergent dynamical sectors with $P_x$ conservation so that the eigenvalues in different sectors do not feel each other and (ii) persistence of local memory starting from certain staggered CDW initial states, which happen to be inert for the effective Hamiltonian with $P_x$ conservation to leading order. This analysis also predicts that the observed `transition' must become a crossover at large sizes once dipole is not strictly conserved --- although various `additional' ingredients in the models in Refs.~\cite{refaelbloch, pollmannstark} such as onsite disorder and non-linearities in the tilt might preclude the eventual thermalization. As an example, while the bare disorder strength in the model in \cite{refaelbloch} looks weak compared to the bare hopping, it may be sizeable compared to the \emph{effective} dipole-conserving hopping and hence lead to MBL via more conventional routes.

Next, we turn to higher dimensions, and see how one can achieve conservation of all components of dipole moment. 
In two or higher dimensions, a tilted potential will result in (prethermal) conservation of only one component of dipole moment if the field/tilt is aligned with a lattice axis ($\hat x$). 
Meanwhile relaxation in directions orthogonal to $\hat x$, corresponding to motion along an equipotential surface, will not be inhibited by the applied field. 

Instead, if the field is applied at an angle $\theta$ with respect to the $\hat x$ axis then it has projections along all the different lattice axes and could potentially engineer long-lived conservation of all components of dipole moment if $F \gg J$. Specializing to 2d, consider $V = F \sum_{\mathbf{r}} \cos(\theta) r_x n_{\mathbf{r}} + \sin(\theta) r_y n_{\mathbf{r}}$. Now, an important point is that if $F_y/F_x = \tan(\theta)$ is rational, then `flat' equipotential lines orthogonal to the tilt direction pass directly through lattice sites and the system can once can again relax along these directions. If $\tan(\theta + \pi/2) = p/q$ then sites along the equipotentials are connected to each other at $O(p+q)$ in the bare nearest-neighbor hopping, giving an effective hopping along the equipotential line $J_{\rm eff} = J (J/F)^{p+q}$, which sets the time-scale for relaxation (the factor of $F$ in the denominator comes from the component of the bare hopping that is against the strong field). Note however that in a purely non-interacting model, there will be a cancellation between `uphill' and `downhill' virtual states, such that the presence of a non-zero interaction is {\it essential} to obtain relaxation along the equipotentials. Another way to see this is to note that for the non-interacting model, the problem is {\it separable} into effectively one dimensional problems along $\hat x$ and $\hat y$ \footnote{we are grateful to Alan Morningstar for this observation}.  

On the other hand, if $\tan(\theta)$ is irrational, there will still be `equipotential surfaces,' although these will not contain more than one lattice site (Fig.~\ref{fig:realize}b), so that in this case one \emph{can} get (prethermal) $P_x, P_y$ conservation along both lattice directions. If we pick a strip of some O(1) width $\epsilon$ about the flat equipotential line, then lattice sites within this strip see a \emph{quasiperiodic} potential set by $F \delta x$, where $\delta x$ is the displacement of the target site from the true equipotential surface. In the interacting problem, two sites in this strip will be `resonant' if the effective hopping/interaction matrix between them exceeds the potential energy difference. This sets a new relaxation timescale $t_c'$ for hopping between distinct lattice sites in an equipotential shell, beyond which it will become apparent that only one component of dipole moment is conserved. If this scale exceeds the prethermal scale $t_*$, then it is irrelevant for the dynamics in which case we expect to find exp(volume) product states that will be eigenstates of the dynamics, up to the prethermal timescale $t_*$. The incommensurate potential may also lead to quasiperiodic MBL~\cite{IyerQP, KhemaniCPQP} along the flat direction for strong enough tilt, in which case only $t_*$ will be relevant and both $P_x, P_y$ look conserved for this time. We note that these arguments are straightforwardly rigorizable using standard prethermalization analyses, and indeed have already been rigorized, following the original posting of this work, in \cite{elseho}.

Finally, let us now discuss how one may generate conservation of quadrupole moment in $d=2$. Quadrupole moment couples to the gradient of the electric field
Thus, the addition of a scalar potential of the form $V(x,y) = F (A x^2 + B y^2 + C x y)$ will, for $O(1)$ coefficients $A,B,C$ and sufficently large $F$, cause the spectrum to split into symmetry sectors labelled by quadrupole moment, again modulo the same considerations as before on prethermalization. However, a scalar potential of this form may be rewritten (at least for $AB>C^2$) simply as $\tilde A (x')^2 + \tilde B (y')^2$, where the $x'$ and $y'$ axes are rotated with respect to the lattice. 
This may simply be recognized as the potential for a {\it harmonic trap}, with inequivalent trap frequencies along the $x'$ and $y'$ directions which is easily realized in experiments (Fig.~\ref{fig:realize}c). 
However, it is not presently clear how to establish conservation of both components of dipole moment \emph{and} quadrupole conservation, in an infinite system. Naively we would think to do this via the addition of a linear component to the potential at an irrational direction with respect to the lattice vectors, which simply shift the trap center along the irrational direction. 
However, a small region far from the trap center will locally only `see' an approximately uniform potential tilt, which along certain `far field' directions will be aligned with lattice axes, leading to conservation of only one component of dipole moment but not both. In a finite size experimental system it may be possible by judicious choice of parameters to evade this issue.

 \section{Discussion and conclusions}
 \label{sec: discussion}
 We have shown how a a finite number of conservation laws can provably `shatter' Hilbert space into a huge number of emergent dynamical subsectors, leading to the emergence of exponentially large localized subspaces in which the localization is robust to temporal noise, does not require disorder, and is characterized by state dependent emergent local integrals of motion. This is in sharp contrast to conventional wisdom which holds that ergodicity breaking requires infinitely many exact or emergent conservation laws. 
 
 The shattering leads to the co-existence, within a particular symmetry sector, of both high and low entanglement states similar to systems with many-body scars. Moreover the unitary operators generating the dynamics may be chosen {\it randomly}, as long as they satisfy the conservation laws, so the model is not at all fine tuned. The key results have been shown to be robust for {\it any} finite gate size, and in any spatial dimension (on hypercubic lattices). While much of our analysis focuses for convenience on dynamics in which the conservation laws function as `hard' constraints that cannot be violated, our results obviously apply also to energy conserving {\it Hamiltonian} dynamics, and can be straightforwardly generalized to settings where the constraints are `soft' (i.e. the conservation laws can be weakly violated). Indeed we specifically discuss both generalizations in the section on physical realizations.
 
  We have also explained how the requisite conservation laws may be naturally introduced in near term ultracold atom experiments. In experimental realizations, the dynamics is Hamiltonian, and the conservation laws are approximate rather than exact. A key signature of the resulting physics lies in the exquisite sensitivity of the dynamics to the initial conditions. Initital conditions with large overlap on the `localized' subspaces should be {\it exact} eigenstates of the dynamics, upto a prethermal timescale that we have estimated. Meanwhile, alternative initial conditions will relax even on timescales short compared to the prethermal timescale. Importantly, the prethermal timescale is always {\it finite} in the thermodynamic limit, so localization from shattering will manifest experimentally as a {\it prethermal crossover} rather than a true transition, although the two may be difficult to distinguish in finite size systems. Our work explains the origin of the recent numerical observations of Stark/Bloch MBL~\cite{refaelbloch, pollmannstark} in tilted finite-size systems, sharpening how the observed non-ergodicity follows from Hilbert space shattering in a large tilt. 

 While `fractonic' models with conservation laws on multipole moments of charge provide the cleanest realization of Hilbert space shattering, we have also provided examples of not obviously fractonic circuits that exhibit shattering. What physical principles underlie these circuits - beyond the fractonic conservation laws discussed herein - would be an interesting topic for future work. We note that our general construction of circuits exhibiting shattering bears a striking resemblance to cellular automata, a connection that may be worth deeper exploration. We also note that a recent work exploring quantum dynamics of cellular automata demonstrated how one may construct exponentially many eigenstates in which at least some sites display trivial dynamics \cite{SarangBahti}. A preliminary exploration of related phenomena in automaton dynamics has also been discussed in \cite{Iaconis}. 
 
 Of course, the broadest {\it physical} class of theories involving local constraints are gauge theories, and `fractonic' phases are known to be describable as gauge theories of `higher rank' \cite{sub}. It would be interesting to explore the possibility of Hilbert space shattering in gauge theories more generally, to clarify whether there are other types of gauge theories (beyond the `fractonic' ones discussed herein) which exhibit such shattering. This may also connect to recent works on ergodicity breaking in gauge theories \cite{NS, DS, konik, lerose, ichinose}.

More generally, this work represents an important addition to the possible classes of many-body quantum dynamics by furnishing a class of models where the dynamics is provably mixed, rather than being either strictly localized or strictly thermalizing for all initial states. Understanding the approach to thermalization for states that do thermalize, and the new classes of dynamical `transitions' between thermalizing and localizing behavior represent important directions for future research.

{\it Note added:} While two of us were finalizing an early version of our manuscript \cite{kn2019}, we learned about related work by P. Sala, T. Rakovszky, R. Verresen, M. Knap and F. Pollmann which appeared in the same arXiv posting \cite{munich}. The results of \cite{munich} have substantial overlap with the discussion in Section \ref{sec: analyze}, and where our results overlap, they agree. 

\section*{Acknowledgments}
We would like to thank Waseem Bakr, Anushya Chandran, David Huse, Jason Iaconis, Michael Knap, Chris Laumann, Alan Morningstar, Sanjay Moudgalya, Shriya Pai, Frank Pollmann, Abhinav Prem, Michael Pretko, Tibor Rakovszky, Pablo Sala, Shivaji Sondhi, Sagar Vijay, and Peter Zoller for useful discussions. We thank Sarang Gopalakrishnan, Alan Morningstar, and Romain Vasseur for feedback on the manuscript.   RN is supported by the Air Force Office of Scientific Research under grant number FA9550-17-1-0183 and by the Alfred P. Sloan Foundation through a Sloan Research Fellowship. This work was supported with funding from the Defense Advanced Research Projects Agency (DARPA) via the DRINQS program (VK). The views, opinions and/or findings expressed are those of the authors and should not be interpreted as representing the official views or policies of the Department of Defense or the U.S. Government. The work of MH is supported by the U.S. Department of Energy, Office of Science, Basic Energy Sciences (BES) under Award number DE-SC0014415. RN and VK also acknowledge the hospitality of the KITP, where part of this work was conducted, during a visit to the program ``Dynamics of Quantum Information''. The KITP is supported in part by the National Science Foundation under Grant No. NSF PHY-1748958.

\appendix

\begin{figure}
\centering
\includegraphics[width=\columnwidth]{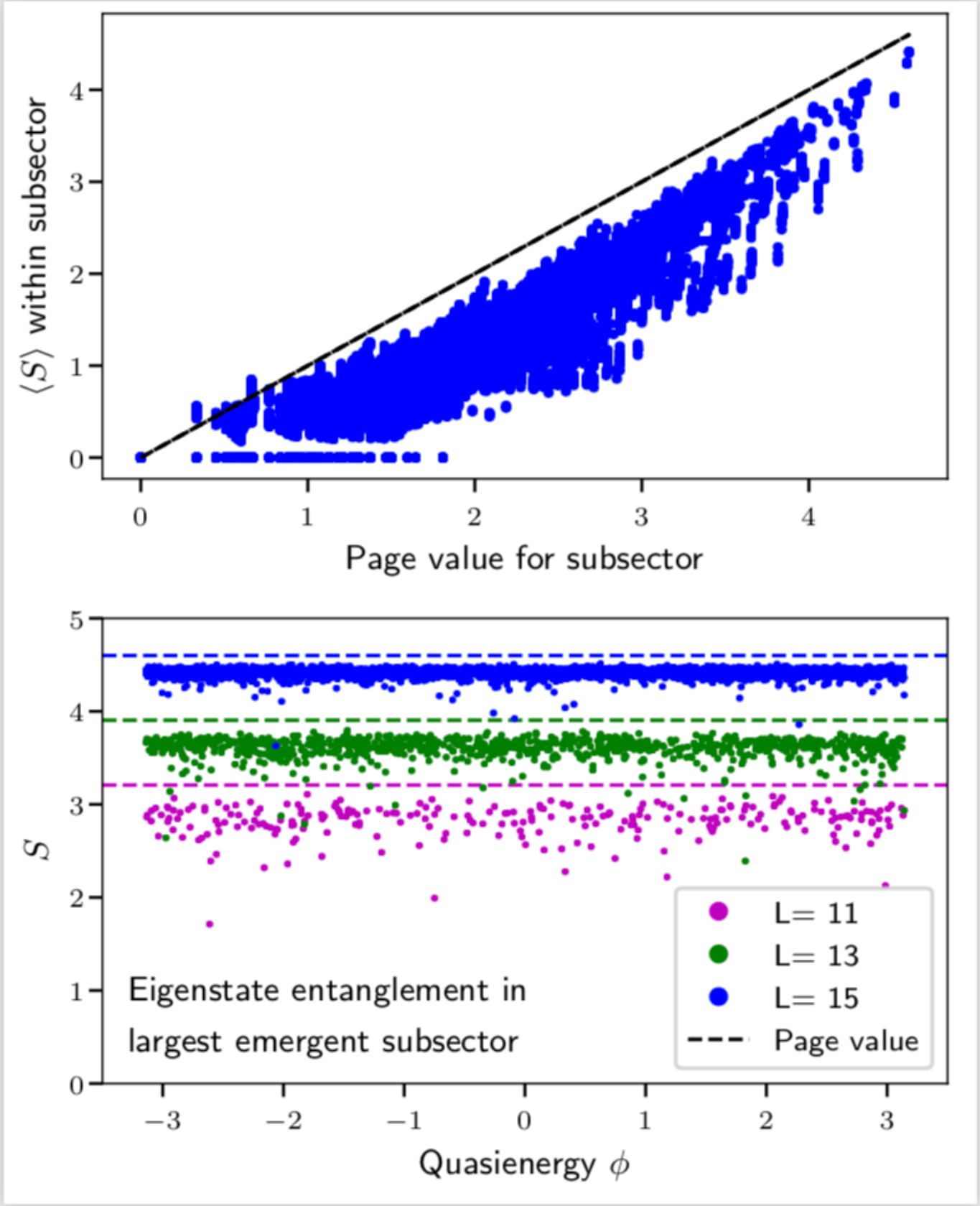}
\caption{\label{subsectors} (a) A plot showing the average bipartite entanglement entropy of eigenstates as a function of subsector size. The `Page' value would be the `thermal' entanglement entropy for a subsector of this size. Data is for $L=15$, three site gates, and open boundary conditions, and all eigenstates with $Q=0,1$ are considered. Note that while the eigenstate entanglement broadly tracks the appropriate Page value for the subsector, there is still a wide distribution, with many eigenstates having significantly subthermal half chain entanglement,  including eigenstates with strictly zero entanglement (`perfect scars') in subsectors that do not exhibit trivial dynamics. This is related to the physics of bottlenecks described in Sec.\ref{bottlenecks}. (b) Entanglement entropy of individual eigenstates within the {\it largest} emergent subsector, plotted as a function of Floquet quasienergy $\phi$. Now the eigenstates do have entanglement close to the `thermal' (Page) value, and this agreement gets better as system size is increased. \label{EE2}}
\end{figure}

\section{Bottlenecks and shielding}
\label{bottlenecks}
In this section, we provide some particular examples of the shielding behavior described in Section~\ref{sec: analyze}, and also discuss a special feature of the fractonic circuit with three site gates, namely the existence of local integrals of motion that can act as `bottlenecks'  \emph{regardless of what larger state they are embedded into}. 

Let us begin with shielding of active regions. 
A simple example for a circuit with three site gates is provided by a configuration of the form $|\cdots 0+0 \cdots \rangle$, where in each case the $\cdots $ denote inert configurations (such as the ones constructed in the previous subsection) ending with a $++$ next to the non-trivial block. 
Applying the allowed $(Q,P)$ conserving moves (Fig.~\ref{fig:floquetCircuit}) readily shows that such a state has non-trivial dynamics only over three sites in real space, and has Hilbert space dimension two. The total charge within this restricted region of real space is then independently a local integral of the motion, even though the circuit is {\it in principle} allowed to spatially move charge. Importantly, this local integral of motion is {\it state dependent} - a single charge immersed in a sea of zeros can move freely by emitting dipoles, whereas a charge blockaded on both sides by inert configurations ending in $++$ cannot leave a restricted region of real space. Multiple analogous ``active" blocks with locally non-trivial dynamics may trivially be introduced into an otherwise `inert' background, each block ``shielded" by $++$ on either end. The size of the active blocks may also be varied in size. Such constructions manifestly exist for any finite gate size, since there is always a localized subspace into which finite non-trivial blocks may be embedded, with appropriate shielding (\emph{cf.} Fig.~\ref{fig:inert}). For example, for a circuit with four site gates, $+++$ would suffice to `shield' a $0+0$ region. These are not the only examples (e.g. all charges could be reversed), but they suffice to make the point that non-trivial blocks can always be embedded into otherwise inert regions.

Next, we turn to bottlenecks, which are motifs that `cut' the system in two, regardless of the state they are embedded in. 
A simple example of such a bottleneck, again for range three gates, is provided by a local pattern of the form $++++$ (or the charged reversed version).  If such a (finite size) pattern is embedded into a larger state that is non-trivial everywhere to the left and the right, then the outer two $+$ charges can move away (by absorbing dipoles), but crucially these `outer' charges perfectly screen the inner charges from dipoles that could make them move. The inner $+$ charges will always be adjacent either to another $+$ charge, or to a $0$, and thus any three-site gate acting on or across the two inner charges must necessarily be trivial (pure phase). As a result, the inner two charges are perfectly localized {\it regardless} of what larger state they are embedded into, and act as a `bottleneck' that cuts the chain in two. The two halves can then be separately labeled by values of charge and dipole moment that are conserved in each half. Likewise, the presence of these bottlenecks at multiple locations can break up the chain into effectively much smaller segments, and the charge and dipole moment of each segment is separately conserved. 

On the other hand, with longer range gates, there is no finite sized motif that can `cut' the chain if embedded into an infinitely large active region. This is easiest to see if one simply embeds the finite sized motif into a sea of zeros. Suppose the left-most site of the `inert/shielding' motif is $+$ (the argument proceeds analogously if it were $-$). One may then create $-++-$ quadrupoles out of the sea of zeros, move off the $-+$ dipole to spatial infinity, and shoot the $+-$ dipole at our motif, causing the leftmost charge to move left one unit. By iterating this process, one can move the leftmost charge of the motif away to spatial infinity, leaving us with a motif reduced in size by one unit, immersed in an infinite sea of zeros (with various charges accumulated at spatial infinity). One may then repeat the process and thus `peel away' the motif one charge at a time. Accordingly, there is no finite sized motif that can `cut' the chain if embedded into an infinite active region, with longer range gates. Of course, a {\it finite} sized active region can always be contained by suitably chosen shielding regions (as discussed in Sec.\ref{larger}) and so even with longer range gates we can have patterns of finite sized `active' and `inert' regions, separated by suitably chosen shielding regions. 

\section{Implications of shattering and shielding for entanglement of eigenstates}
\label{app:entanglement}
In this appendix we re-examine our results on the mid-cut entanglement entropy of eigenstates, armed with our understanding of Hilbert space fracture. The first point is that there are emergent dynamical subsectors of varying sizes even within a single $(Q,P)$ sector, and the ``thermal" value for eigenstates in a given dynamical subsector will be controlled by the size of the subsector \cite{DonPage}. For instance, in the extreme case of strictly inert states, the eigenstate entanglement will be exactly zero. More generally, the subsectors of various sizes naturally lead to a broad distribution of high and low entanglement states -- ranging from area to volume law --  within a single extensive symmetry sector, as was observed in Fig.~\ref{EE}. 

To examine this more quantitatively, in Fig~\ref{EE2}(a), we plot the average entanglement entropy of each emergent dynamical subsector against the thermal (Page) value for that subsector in a system of length $L=15$ with three-site gates. We consider all eigenstates in all subsectors in the $Q=\{0,1\}$ sectors (with all possible $P$ values). The data is averaged over $~100$ independent circuit realizations. 
The Page value is computed by explicitly examining the $S^z$ basis states that span a given subsector, and using these to extract $\mathcal{D}_L$ and $\mathcal{D}_R$, the dimension of the Hilbert spaces in the left and right halves of the chain for that subsector. Because of the constraints, these depend on the exact basis states that form the subspace and could be different for different subsectors of the same size. Because some of the subsector sizes are very small, we use the exact expression for the Page value~\cite{DonPage} $S_{\rm Page} = \sum_{k={n+1}}^{mn} \frac{1}{k} - \frac{m-1}{2n}$, where $m = \rm{min}[\mathcal{D}_L, \mathcal{D}_R]$ and $n = \rm{max}[\mathcal{D}_L, \mathcal{D}_R]$; this reduces to the more familiar form $S_{\rm Page}\sim \log(n) - \frac{m}{2n}$ for $1 \ll m\leq n$. 

A priori one might have thought that the existence of these multiple subsectors with a broad distribution of sizes would be sufficient to explain the co-existence of high and low entanglement states within a symmetry sector. Indeed, the eigenstate entanglement does broadly track the Page value for the appropriate subsector, as shown in Fig~\ref{EE2}(a). However, the figure also shows the existence of a broad distribution of entanglement entropies {\it even after} resolving by subsector size. There even exist states with strictly zero entanglement in subsectors with dimension greater than one. Thus, the `shattering' of Hilbert space is part of the explanation for the broad distribution of entanglement entropies, but it is not the whole picture.  

This brings us to our second point --  a key part of the explanation for the broad distribution of entanglement entropies, even after resolving by subsector size, is the bottleneck/shielding phenomenon discussed in Secs.~\ref{bottlenecks}, \ref{larger}. In particular, the states with zero entanglement entropy (which are not in the strictly localized subspace) have been explicitly verified to contain a `bottleneck' motif at the midpoint of the chain, which prevents development of any entanglement across this motif, which happens to overlap the entanglement cut. The existence of such `bottleneck' motifs at positions away from the entanglement cut is also at least partially responsible for the existence of a broad distribution of entanglement entropies, even after resolving by subsector size, since the effective number of entangling degrees of freedom get reduced when the chain is `cut'. More generally, the entanglement entropy is bounded by the Hilbert space dimension of the active region that straddles the cut, and this can be much less than the size of the subsector in which the state lives, if the state consists of disconnected active regions. This discussion highlights that not only is there strong state-to-state variation in the entanglement properties of eigenstates, there is also a strong variation across spatial locations of the entanglement within a given state. 

Finally, we note that the entanglement entropy in the subsector of largest size \emph{does} appear to well approximate the thermal Page value, and this agreement gets better with increasing system size (Fig.~\ref{EE2}(b)).

\section{Implications of shattering for dynamics}
\label{app:dynamics}
In this appendix we discuss at length the implications of shattering for dynamics starting from different initial states. Note that while we have proven the existence of an exponentially large localized subspace, this subspace is still a measure zero fraction of the entire Hilbert space in the thermodynamic limit. While initial conditions that have high overlap with this localized subspace will clearly exhibit localization, initial conditions chosen {\it randomly} in Hilbert space will have vanishing overlap with the localized subspace. We now discuss the implications of Hilbert space shattering for the dynamics from random initial conditions. 

Dynamics from random initial conditions is expected to be highly sensitive to the degree of shattering. In Fig.\ref{longrangegates}(c) we examine what fraction of the states in a symmetry sector are contained in the emergent subsector of largest size. For three site gates, the largest emergent subsector is observed to contain a vanishing fraction of the states in the thermodynamic limit, consistent with our analytic estimates. (Recall that the largest subsector contained $\sim 2^L$ states, whereas the Hilbert space dimension is $3^L$). In contrast, for longer range gates a non-zero fraction (almost exactly equal to one) of the Hilbert space is contained in the emergent subsector of largest size, and this does not change with changing system size. 

\begin{figure}
\centering
\includegraphics[width=\columnwidth]{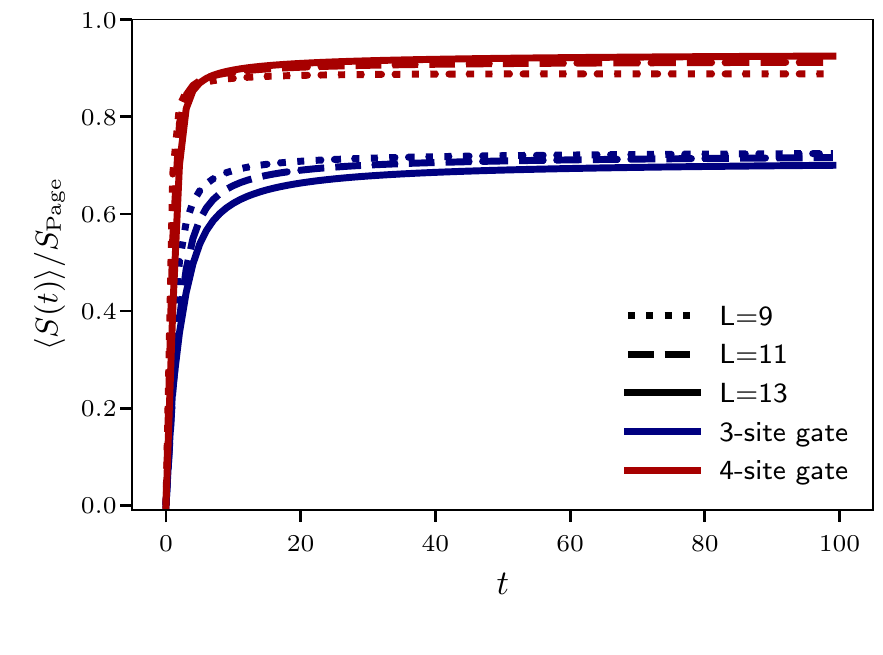}
\caption{Figure showing the dynamics of entanglement starting from a random product state (not in the $z$ basis). Entanglement entropy is given as a ratio of the `Page' value for a random product state in a Hilbert space of dimension $3^L$. For three site gates, the entanglement entropy saturates to well below its Page value, consistent with expectations given the strong fracturing of Hilbert space. For four site gates, the saturating value of entanglement entropy is much closer to the Page value, with a slow upward drift with increasing system size. \label{EEDynamics}}
\end{figure}

These differences may have interesting implications for the dynamics from randomly chosen initial product states (which are not in the $z$ basis and are not confined to any particular symmetry sector).  
For example, for three site gates, the largest subsector of Hilbert space has dimension $\sim 2^L$. The late-time entanglement entropy should therefore be dominated by this subsector and scale as $~L \ln 2$. Meanwhile, the entire Hilbert space has dimension $3^L$, and so the thermal or `Page' entanglement entropy for the full Hilbert space is of order $L \ln 3$. We would therefore expect that for a circuit with three site gates, a random initial condition should exhibit entanglement entropy growth saturating to a value approximately equal to $\frac{\ln 2}{\ln 3} S_{Page} \approx 0.63 S_{Page}.$ However, for a circuit with four site gates, the largest subsector size and the Hilbert space dimension scale similarly, as $3^L$, and one might expect dynamics starting from random initial conditions to lead to entanglement entropy growth saturating close to the Page value. (Note however a potential loophole on this argument - if the states in the largest subsector were made up of disconnected active subregions, then the saturating entropy would be bounded by the Hilbert space dimension of the active subregion straddling the entanglement cut, which could well be less than the Hilbert space dimension of the entire subsector). 

To test this intuition, in Fig.~\ref{EEDynamics}, we show the growth of entanglement entropy for both three and four site gates, starting from an initial condition that is a random product state. Note that a random product state (not in the z basis) is a superposition of multiple  symmetry sectors and subsectors. For three site gates, the entanglement entropy is observed to saturate to a clearly subthermal value of order $0.6 S_{Page}$, consistent with our expectations. Meanwhile, for four site gates the saturation value for the entanglement entropy is clearly higher, much closer to the Page value, with a slow upward drift with increasing system size. Whether the saturating value of entanglement entropy actually reaches the Page value in the thermodynamic limit is not clear from the present numerics. A more extensive investigation of pure state dynamics starting from random initial conditions, and how this depends on gate range, would be an interesting problem for future work.

\section{Localized subspace for fractonic circuit with four site gates}
\label{app:foursitecounting}
\begin{table}
\begin{tabular}{c|c}
\underline{Last three sites of L site chain are} & \underline{Site added can be} \\
+++ & + or 0 or -\\
++0 & + or 0 or -\\
++- & - \\
+0+ & +\\
+0-& - \\
+- -&- \\
0++ & +\\
00+ & +\\
00- & -\\
0- -&- \\
-++ & +\\
-0+ & +\\
-0- & -\\
- -+ & +\\
- - 0 & + or 0 or -\\
- - - & + or 0 or -\\
+00 & - \\
-00 & +
\end{tabular}
\caption{For the fractonic circuit with four site gates, if an inert state in a system of size $L$ has the final three sites in the states shown in the left column, then it remains inert upon addition of another spin if the new spin is in the corresponding state shown in the right column. Note that we have only listed sixteen of the twenty seven possible configurations for the last three spins of the $L$ site chain - the remaining eleven configurations are `dead ends' i.e. there is nothing that can be added that leaves the state inert. \label{four}}
\end{table}

In this Appendix we provide an explicit calculation of the localized subspace for the fractonic circuit with four site gates. In this case the gates are matrices of rank $3^4=81$, with structure as detailed in Table I of \cite{pai2018localization}. Note however that there is a typo in the charge zero block of that table, in that configurations such as $+00-$ and $-00+$ should be inert, whereas $+-+-$ should mix freely with $+0-0$ and $0+0-$, but not with $+00-$. With this typo corrected, we note that in a chain of size $L=4$ there are twenty six trivial states. If a state is inert in an $L$ site system, then the addition of another site will leave it still inert as long as the last three sites of the $L$ site chain and the added site collectively form an inert state of the $L=4$ chain i.e. if the conditions detailed in Table \ref{four} are fulfilled. Note that of the twenty seven possible end states for a chain of length $L$, only eighteen allow the state to remain inert upon addition of another spin - the rest are `dead ends.' This is an important distinction to the circuit with three site gates where there were no dead ends. We can then write a recursion relation for the eighteen `live' configurations only, and it takes the form of the rank eighteen matrix equation given below.

\begin{widetext}
\begin{eqnarray}
\left( \begin{array}{c} N_{+++}\\ N_{++0}\\ N_{++-}\\ N_{+0+}\\N_{+0-} \\N_{+--} \\ N_{0++} \\ N_{00+} \\ N_{00-} \\ N_{0--} \\ N_{-++} \\ N_{-0+} \\ N_{-0-} \\ N_{- -+}\\N_{- -0}\\N_{- - -} \\ N_{+00} \\ N_{-00}
\end{array}
\right)_{L+1} = \left(\begin{array}{cccccccccccccccccc} 1 & 0 & 0 & 0 & 0 & 0 & 1 & 0 & 0 & 0 & 1 & 0 & 0 & 0 & 0 & 0 & 0 & 0 \\ 1 & 0 & 0 & 0 & 0 & 0 & 0 & 0 & 0 & 0 & 0 & 0 & 0 & 0 & 0 & 0 & 0 & 0 \\ 1 & 0 & 0 & 0 & 0 & 0 & 0 & 0 & 0 & 0 & 0 & 0 & 0 & 0 & 0 & 0 & 0 & 0  \\ 1 & 0 & 0 & 0 & 0 & 0 & 0 & 0 & 0 & 0 & 0 & 0 & 0 & 0 & 0 & 0 & 0 & 0 \\ 1 & 0 & 0 & 0 & 0 & 0 & 0 & 0 & 0 & 0 & 0 & 0 & 0 & 0 & 0 & 0 & 0 & 0 \\ 0 & 0 & 1 & 0 & 0 & 0 & 0 & 0 & 0 & 0 & 0 & 0 & 0 & 0 & 0 & 0 & 0 & 0 \\ 0 & 0 & 0 & 1 & 0 & 0 & 0 & 1 & 0 & 0 & 0 & 1 & 0 & 0 & 0 & 0 & 0 & 0 \\ 0 & 0 & 0 & 0 & 0 & 0 & 0 & 0 & 0 & 0 & 0 & 0 & 0 & 0 & 0 & 0 & 0 & 1\\ 0 & 0 & 0 & 0 & 0 & 0 & 0 & 0 & 0 & 0 & 0 & 0 & 0 & 0 & 0 & 0 & 1 & 0 \\ 0 & 0 & 0 & 0 & 1 & 0 & 0 & 0 & 1 & 0 & 0 & 0 & 1 & 0 & 0 & 0 & 0 & 0 \\ 0 & 0 & 0 & 0 & 0 & 0 & 0 & 0 &0 & 0 & 0 & 0 & 0 & 1 & 0 & 0 & 0 & 0 \\  0 & 0 & 0 & 0 & 0 & 0 & 0 & 0 &0 & 0 & 0 & 0 & 0 & 0 & 1 & 0 & 0 & 0 \\ 0 & 0 & 0 & 0 & 0 & 0 & 0 & 0 &0 & 0 & 0 & 0 & 0 & 0 & 1 & 0 & 0 & 0 \\ 0 & 0 & 0 & 0 & 0 & 0 & 0 & 0 &0 & 0 & 0 & 0 & 0 & 0 & 0 & 1 & 0 & 0 \\ 0 & 0 & 0 & 0 & 0 & 0 & 0 & 0 &0 & 0 & 0 & 0 & 0 & 0 & 0 & 1 & 0 & 0 \\ 0 & 0 & 0 & 0 & 0 & 1 & 0 & 0 & 0 & 1 & 0 & 0 & 0 & 0 & 0 & 1 & 0 & 0 \\ 0 & 1 & 0 & 0 & 0 & 0 & 0 & 0 & 0 & 0 & 0 & 0 & 0 & 0 & 0 & 0 & 0 & 0 \\  0 & 0 & 0 & 0 & 0 & 0 & 0 & 0 & 0 & 0 & 0 & 0 & 0 & 0 & 1 & 0 & 0 & 0  \end{array} \right)
\left( \begin{array}{c} N_{+++}\\ N_{++0}\\ N_{++-}\\ N_{+0+}\\N_{+0-} \\N_{+--} \\ N_{0++} \\ N_{00+} \\ N_{00-} \\ N_{0--} \\ N_{-++} \\ N_{-0+} \\ N_{-0-} \\ N_{- -+}\\N_{- -0}\\N_{- - -} \\ N_{+00} \\ N_{-00}
\end{array}
\right)_{L} \nonumber
\end{eqnarray}
The largest eigenvalue of the above matrix has magnitude $1.8$, leading us to conclude that the dimension of the localized subspace grows asymptotically as $1.8^L$, in agreement with Figure~\ref{longrangegates}(a) and again, faster than the lower bound of $2^{L/4} \sim 1.2^L$. Similar analyses may be carried through for any finite range of gates in the fractonic circuit, but the analysis rapidly becomes tedious and so we do not pursue it here. 
\end{widetext}

\bibliography{library}

\end{document}